\documentclass[usenatbib]{mnras}

\usepackage{newtxtext,newtxmath}
\usepackage[T1]{fontenc}
\usepackage{ae,aecompl}

\usepackage{graphicx}	% Including figure files
\usepackage{amsmath}	% Advanced maths commands
\usepackage{amssymb}	% Extra maths symbols
\usepackage{amstext}
\usepackage[usenames]{xcolor}

\newcommand{\Msun}{\,M_{\odot}}

\newcommand{\Ms}{M_\star}
\newcommand{\epsff}{\epsilon_{\mathrm{ff}}}
\newcommand{\ZH}{\mathrm{[Z/H]}}
\newcommand{\feh}{\mathrm{[Fe/H]}}
\newcommand{\lambdam}{\lambda_{\mathrm{m}}}

\title[Mass and metallicity distribution of GCs]{Star cluster formation in cosmological simulations -- III. Dynamical and chemical evolution}

\author[Li et al.]{
Hui Li$^{1}$\thanks{E-mail: hliastro@mit.edu}\href{https://orcid.org/0000-0002-1253-2763}{\includegraphics[scale=0.6]{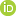}} and
Oleg Y. Gnedin$^{2}$\href{https://orcid.org/0000-0001-9852-9954}{\includegraphics[scale=0.6]{orcid.png}}
\\
$^{1}$Department of Physics, Kavli Institute for Astrophysics and Space Research, MIT, Cambridge, MA 02139, USA\\
$^{2}$Department of Astronomy, University of Michigan, Ann Arbor, MI 48109, USA
}

\date{Accepted XXX. Received YYY; in original form ZZZ}

\pubyear{2018}

\begin{document}
\label{firstpage}
\pagerange{\pageref{firstpage}--\pageref{lastpage}}
\maketitle

\begin{abstract}
In previous papers of this series, we developed a new algorithm for modelling the formation of star clusters in galaxy formation simulations. Here we investigate how dissolution of bound star clusters affects the shape of the cluster mass function and the metallicity distribution of surviving clusters. Cluster evolution includes the loss of stars that become unbound due to tidal disruption as well as mass-loss due to stellar evolution. We calculate the tidal tensor along cluster trajectories and use it to estimate the instantaneous mass-loss rate. The typical tidal tensor exhibits large variations on a time-scale of $\sim100$~Myr, with maximum eigenvalue of $10^7$~Gyr$^{-2}$, and median value of $10^4$~Gyr$^{-2}$ for the first Gyr after cluster formation. As a result of dynamical disruption, at the final available output of our simulations at redshift $z\approx1.5$, the cluster mass function has an approximately log-normal shape peaked at $\sim10^{4.3}\Msun$. Extrapolation of the disruption to $z=0$ results in too many low-mass clusters compared to the observed Galactic globular clusters (GCs).
Over 70\% of GC candidates are completely disrupted before the present; only 10\% of the total GC candidate mass remains in surviving clusters. The total mass of surviving clusters at $z=0$ varies from run to run in the range $(2-6)\times10^7\Msun$, consistent with the observed mass of GC systems in Milky Way-sized galaxies. The metallicity distributions of all massive star clusters and of the surviving GCs have similar shapes but different normalization because of cluster disruption. The model produces a larger fraction of very metal-poor clusters than observed. A robust prediction of the model is the age-metallicity relation, in which metal-rich clusters are systematically younger than metal-poor clusters by up to 3~Gyr.
\end{abstract}

\begin{keywords}
galaxies: evolution --- galaxies: formation --- galaxies: star clusters: general ---  globular clusters: general --- methods: numerical 
\end{keywords}

%%%%%%%%%%%%%%%%%%%%%%%%%%%%%%%%%%%%%%%%%%%%%%%%%%

%%%%%%%%%%%%%%%%% BODY OF PAPER %%%%%%%%%%%%%%%%%%

\section{Introduction}\label{sec:intro}

All galaxies with stellar mass larger than $10^9\Msun$ in the local Universe host globular cluster (GC) systems \citep[][]{brodie_strader06}. Recently revealed tight correlation between the total mass of a GC system and the host galaxy halo mass \citep{peng_etal06, jordan_etal07,blakeslee_etal97, spitler_forbes09,georgiev_etal10, harris_etal13, harris_etal15} suggests that GC systems in some way preserve the memory of the assembly history of their host galaxies. However, due to lack of comprehensive understanding of the co-evolution of GCs and their host galaxies, making use of observations of extragalactic GC systems to reconstruct galaxy evolution is still far from feasible.

One major uncertainty that hinders our understanding of the origin of GCs is their dynamical evolution across cosmic time. In the past decade, inspired by the discoveries of a large number of young massive star clusters that share similar mass and size to GCs \citep[see][for a review]{portegies_zwart_etal10}, it has been generally agreed that the observed log-normal mass function of GCs at present is a relic of a truncated power-law initial mass function of young clusters formed at high redshift \citep[e.g.][]{fall_zhang01, kroupa_boily02, vesperini_etal03, prieto_gnedin08, mclaughlin_fall08, kruijssen09, muratov_gnedin10, li_gnedin14, webb_leigh15, carlberg18}. The dynamical evolution of massive clusters is governed by a combination of stellar evolution, internal two-body relaxation, external tidal truncation, and tidal shocks. Although the mass-loss via stellar evolution for single stellar populations \citep[e.g.][]{leitherer_etal99} and internal two-body relaxation for isolated clusters \citep{spitzer87} are well studied, the tidal disruption process has not been modelled sufficiently in the full context of hierarchical galaxy formation.
Previous works on tidal disruption usually assume an idealized analytical tidal field experienced by clusters along circular or elliptical orbits in the host galaxy \citep[e.g.][]{takahashi_pz00, baumgardt_makino03, prieto_gnedin08, penarrubia_etal09, rieder_etal13, webb_etal14, rossi_etal16}, see a recent review by \citet{renaud18}. This simplified picture ignores the structure and granularity of the gravitational potential, and therefore cannot capture tidal interactions with the dense disc, giant molecular clouds (GMCs), and other star clusters, all of which are believed to make important contributions to disruption.

Recent efforts have been made to obtain more realistic tidal fields from hydrodynamic simulations of galaxy formation. For example, \citet{renaud_etal17} post-processed a cosmological simulation of a Milky Way-sized galaxy and found that the average strength of the tidal field around stellar particles increases towards lower redshift. More recently, \citet{pfeffer_etal18} introduced the E-MOSAICS simulation, based on the EAGLE galaxy formation model \citep{schaye_etal15} with added runtime modelling of cluster formation and disruption. One limitation of the above approaches is that they do not model  the creation of individual clusters in star-forming regions. Another, more crucial issue is the spatial resolution of these simulations which is too coarse to resolve the dense structure in the galactic disc. Therefore, the strength of the tidal field can be significantly underestimated, especially when clusters are still interacting with dense GMCs shortly after their formation \citep[e.g.][]{gieles06}. 

Another important property of GC systems is their metallicity distribution. Observations of Galactic and extra-galactic GCs often reveal a bimodal colour distribution, which can translate to a bimodal metallicity distribution \citep{zepf_ashman93,whitmore_etal95,larsen_etal01,peng_etal06}. Interestingly, the largest GC systems in brightest cluster galaxies show a unimodal distribution with broad metallicity range \citep{harris_etal16, harris_etal17b}. Understanding what processes shape the metallicity distribution can shed light on the mass assembly and chemical enrichment history of massive galaxies.

In \citet[][hereafter, Paper~I and Paper~II, respectively]{li_etal17,li_etal18} we developed a new algorithm for modelling the formation of star clusters in cosmological hydrodynamic simulations. The mass of individual cluster particles is determined by the balance between continuous gas accretion from the ambient medium and the energy and momentum feedback from the cluster's own stars. The final mass is set self-consistently and represents the mass of a bound star cluster emerged from its natal GMC. We performed a suite of cosmological simulations with very high spatial resolution, $3-6$~pc, to resolve dense parts of typical GMCs. These simulations use strong stellar feedback and match the stellar masses and star formation rates of high-redshift galaxies expected from the abundance matching method. The initial mass function of model star clusters is consistent with the observations of young star clusters in the nearby galaxies. This implementation of cluster formation provides us with an opportunity to study the dynamical and chemical evolution of clusters in realistic galactic environments. The extremely high spatial and temporal resolutions of these simulations allow us to accurately calculate the tidal field around individual star clusters throughout their evolution.

In this third paper of the series, we introduce a new algorithm to calculate the tidal field along the orbit of each cluster and estimate the tidal disruption rate at each global time-step of the simulation. This is the first time that self-consistent star cluster formation and disruption processes are calculated simultaneously in the runtime of cosmological simulations with such high spatial and temporal resolution. In the course of the simulation, the metallicity of model clusters is tracked by metal enrichment from both Type II and Ia supernovae. We show that tidal disruption changes the cluster mass function from the initial Schechter-like to a log-normal shape, which in turn changes the metallicity distribution of the surviving clusters.

We first recap some key features of the star cluster formation prescription that is developed in Paper I and II and the suite of simulations that are used in this paper in \autoref{sec:simulations}. In \autoref{sec:dyn}, we describe the algorithm to calculate the tidal field and mass-loss, and present the orbit of one representative cluster, the time variation of the tidal fields, and the evolution of the cluster mass function. In \autoref{sec:metal}, we describe the chemical enrichment and metal accretion process for model clusters, and examine their metal spread and metallicity distribution. We discuss the implications of our results for GC formation in \autoref{sec:discussion} and summarize our conclusions in \autoref{sec:summary}.

\section{Simulation Suite}\label{sec:simulations}

Detailed description of the simulation setup and implementation of the cluster formation algorithm are presented in Papers I and II. Here we briefly recap some key features and parameters of the simulations we use in this paper.

The simulations are performed with the Eulerian gasdynamics and N-body Adaptive Refinement Tree code \citep[ART,][]{kravtsov_etal97,kravtsov99,rudd_etal08}. They are run from cosmological initial conditions that contain a main dark matter halo of total mass $M_{200}\approx 10^{12}\Msun$ at $z=0$, within a periodic box of 4 comoving Mpc in size. The simulations start with a $128^3$ root grid with the mass of dark matter particles $m_{\rm DM}=1.05\times 10^6\Msun$. We use adaptive mesh refinement technique to achieve high dynamical range of the spatial resolution. At very high redshift we allow a maximum of 9 additional levels of refinement. Then we add the 10th, 11th, and 12th additional levels at $z=9$, 4, and 1.5, respectively, to maintain a roughly fixed physical size of the highest-refinement cells. The smallest cell size remains in the range between 3 and 6 pc, comparable to the typical size of star-forming regions in nearby galaxies.

In Paper I, we introduced a new star formation prescription in the ART code: continuous cluster formation. In this prescription, each star particle represents an individual star cluster and is created at the local density peak of the molecular gas. After creation, the cluster particle grows its mass via gas accretion within a spherical region of the fixed physical size $R_{\rm GMC}=5$~pc. The accretion rate is determined by the local $\rm H_2$ density of cells within the sphere:
\begin{equation}
  \dot{M} = \frac{\epsff}{\tau_{\rm ff}}\, \sum_{\mathrm{cell}}{f_{\rm GMC}\, V_{\mathrm{cell}}\, f_{\rm H_2}\, \rho_{\rm gas}},
  \label{eq:sfr}
\end{equation}
where $V_{\mathrm{cell}}$ is the volume of each neighbour cell, $f_{\rm GMC}$ is the fraction of $V_{\mathrm{cell}}$ that overlaps with the GMC, $f_{\rm H_2}$ is the mass fraction of hydrogen in molecular phase, $\rho_{\rm gas}$ is the total gas density within a cell, and $\epsff$ is the local star formation efficiency per free-fall time $\tau_{\rm ff}$. As the cluster particle grows its mass, it injects mass, energy, and momentum to the ambient medium via stellar winds, radiative pressure, and supernova explosions. The energy budget for these processes is calibrated using the stellar population synthesis model with \citet{kroupa01} initial stellar mass function. Moreover, we use a recent implementation of the supernova remnant feedback model, in which the thermal, kinetic, and turbulent energy injections are calibrated by high-resolution hydrodynamic simulations in an inhomogeneous turbulent medium \citep{martizzi_etal15}. We also adopt a momentum boosting factor, $f_{\rm boost}$, to take into account the enhancement of momentum feedback from clustering of supernova explosion \citep{gentry_etal17}. We found with $f_{\rm boost}=5$, the star formation history of the main galaxy best matches the abundance matching results. Therefore, we choose $f_{\rm boost}=5$ as the default value for all runs used in this paper.

For our analysis of the cluster mass function and metallicity distribution we use the most advanced simulations described in Paper II, with the updated cluster creation/accretion algorithm and feedback module. Specifically, we consider runs SFE10, SFE50, SFE100, and SFE200, which reached the epoch $z\approx 1.5$\footnote{Most of our simulations did not advance beyond redshift 1.5, because the required time-step became extremely short due to the fast winds launched from the star-forming regions.}. These runs are distinguished by the value of $\epsff=0.1-2.0$. New for this paper, we have continued run SFE200 to redshift $z\approx 0.6$. We use this run to analyse the evolution of orbits and tidal fields around several representative clusters.

\section{Dynamical evolution of star clusters}\label{sec:dyn}
\subsection{Evolution of cluster bound fraction}

After cluster particles are created, they begin to lose mass via a combination of physical processes, such as mass-loss from individual stars due to stellar evolution and escape of unbound stars due to two-body scattering with other stars and variation of the external tidal field. The fraction of the initial cluster particle mass that is gravitationally self-bound is defined as the bound fraction
$$ 
   f_{\rm bound}(t) =  f_{\rm i}\, f_{\rm se}(t)\, f_{\rm dyn}(t). 
$$ 
Here $f_{\rm i}$ is the initial bound fraction accounting for the gas expulsion during cluster formation (defined and calculated in Paper~II), $f_{\rm se}$ accounting for the mass-loss due to stellar evolution, and $f_{\rm dyn}$ accounting for the tidal stripping of stars. From Paper II, we found that $f_{\rm i}$ shows a strong positive correlation with the cluster particle mass and increases systematically with the increase of $\epsff$. As will be seen later, because of this dependence, the mass and metallicity distributions of the surviving clusters show dramatic difference for runs with different $\epsff$.

Our implementation of stellar evolution assumes a time-dependent stellar wind and stellar explosion model that is calibrated by the FSPS population synthesis code \citep{conroy_etal09, conroy_gunn10}. The integrated mass-loss rate of a simple stellar population with a \citet{kroupa01} IMF is described in Figure~2 and Section~2.2.7 of Paper~II.

Independent of stellar evolution, the dynamical destruction of star clusters, $f_{\rm dyn}$, takes into account the escape of unbound stars due to tidal disruption when clusters are orbiting around their host galaxy. Accurate modelling of the dynamical evolution requires collisional N-body or Fokker-Planck approaches, which are beyond the scope of this paper. Instead, we characterise the mass-loss by a tidal disruption time-scale, $t_{\rm tid}$. Following \citet{gnedin_etal14}, the mass-loss rate of the tidal disruption can be expressed as:
\begin{equation} \label{eq:bound-mass}
	\frac{dM}{dt}=-\frac{M}{t_{\rm tid}}.
\end{equation}

In the following sections, we describe how we calculate the disruption rate of clusters under the time-varying tidal fields, and how we update the bound fraction $f_{\rm bound}$ given the disruption rate in the simulation runtime.

\subsection{Tidal field around clusters in realistic galactic environments} \label{sec:disruption-tidalfield}

The strength and orientation of the tidal field can be fully characterised by the tidal tensor. The general formalism of a tidal tensor at position $\mathbf{x}_0$ under a time-varying gravitational potential field $\Phi(\mathbf{x},t)$ is defined as
\begin{equation}
  T_{ij}(\mathbf{x}_0, t) \equiv -\frac{\partial^2\Phi(\mathbf{x},t)}{\partial x_i \, \partial x_j}\Bigr|_{\mathbf{x}=\mathbf{x}_0},
\end{equation}
where $i$ and $j$ are two orthogonal directions in the Cartesian galactocentric coordinate frame.

We estimate the tidal tensors around all fully formed (inactive, age above 15~Myr) clusters in runtime of the simulations at each global time step.\footnote{The length of a global time step varies during the course of the simulation. It is typically around $1-3$~Myr.} Algorithmically, the second-order finite differences of the gravitational potential are calculated across the $3\times3\times3$ cell cube centered on a given cluster. For the diagonal terms, e.g.:
\begin{align*}
  T_{xx} &= -\frac{\partial^2\Phi(x,y,z)}{\partial x^2} \\
& \approx
  -\frac{\Phi(x+L_{\rm cell},y,z)+\Phi(x-L_{\rm cell},y,z)-2\Phi(x,y,z)}{L_{\rm cell}^2},
\end{align*}
where $L_{\rm cell}$ is the cell size. For non-diagonal terms, e.g.:
\begin{align*}
  T_{xy} &= -\frac{\partial^2\Phi(x,y,z)}{\partial x\, \partial y} \\
& \approx
  -\frac{1}{4L_{\rm cell}^2} \left[ \Phi(x+L_{\rm cell},y+L_{\rm cell},z)+\Phi(x-L_{\rm cell},y-L_{\rm cell},z) \right.\\
&\left. \hspace{1.5cm}-\Phi(x+L_{\rm cell},y-L_{\rm cell},z)-\Phi(x-L_{\rm cell},y+L_{\rm cell},z) \right].
\end{align*}
Using the above second-order finite difference scheme, the tidal tensor is estimated over the scale of twice the cell size $2L_{\rm cell}$. Although the spatial resolution of the simulations can reach as high as $3-6$~pc in the densest star-forming regions, after formation clusters typically migrate to much less dense environments where the cells are larger, $L_{\rm cell}\sim 12-24$~pc. This means that the scale of the tidal field calculation is at least 24~pc, which is much larger than the effective radii of young massive star clusters observed in the nearby galaxies \citep[e.g.][]{portegies_zwart_etal10}. This scale is also comparable to the estimates of tidal radii of the Galactic GCs, which are of the order 20-50 pc.
Thus our scheme is appropriate for capturing the tidal field around the model clusters. Moreover, the gravitational potential contributed by dark matter particles is smoothed at the first 6 refinement levels. The grid size at the 6-th level ($\approx488/(1+z)$~pc) is large enough so that the discreteness of the potential by individual dark matter particles does not affect the tidal field estimation.

We then calculate the three eigenvalues of the tidal tensor, $\lambda_1\ge\lambda_2\ge\lambda_3$, which represent the intensity of the tidal field in the direction of the corresponding eigenvectors. The tidal disruption time-scale can then be inferred from these eigenvalues, as we describe below.

\subsection{Disruption time-scales and the bound fraction} \label{sec:disruption-time-scales}

N-body simulation results of \citet{gieles_baumgardt08} in an idealized setup of clusters on circular orbits in a static Galactic potential show that the tidal disruption time-scale depends strongly on the orbital angular frequency, $\Omega_{\rm tid}$:
$$
    t_{\rm tid}(M) \propto M(t)^{2/3}\ \Omega_{\rm tid}^{-1}.
$$
These N-body runs were designed for young open clusters, and therefore included a wide spectrum of stellar mass (a factor of 30) and no stellar evolution. The presence of massive stars speeds up the relaxation process: \citet{gieles_etal10} estimated that the rate of relaxation relative to the equal-mass case scales as the square root of the ratio of maximum to minimum stellar mass. In their 2008 runs the ratio was taken to be a factor of $\sim 4$ higher than observed in GCs, which should lead to a factor of 2 quicker disruption. This shorter disruption time was adopted in \citet{gnedin_etal14}. Indeed, the simulations by \citet{lamers_etal10} that take into account stellar evolution and the corresponding reduction of the mass ratio indicate a longer disruption time-scale, roughly by a factor of 2.5 for clusters of typical initial mass $2\times 10^5\,\Msun$. Such a time-scale is consistent with that derived from observations, see \citet{lamers_etal05}. We are grateful to Mark Gieles for pointing out this discrepancy. Correspondingly, we have revised our normalization of the disruption time-scale as
\begin{equation}
  t_{\rm tid}(M,t) \approx 10~{\rm Gyr} \left(\frac{M(t)}{2\times 10^5\Msun}\right)^{2/3}
    \frac{100 \,\mathrm{Gyr}^{-1}}{\Omega_{\rm tid}(t)}.
  \label{eq:t-tid}
\end{equation}

A generalization for more realistic orbits would have $\Omega_{\rm tid}$ characterise the strength of the actual tidal field around a cluster. One option is to use the dynamical time-scale within the Roche lobe of the cluster:
$$ \Omega_{\rm tid}^2 \approx \frac{1}{3}\lambdam $$
where
$$ \lambdam \equiv \max|\lambda_i|. $$ 
In this case, both pure extensive and compressive tides will contribute to the disruption process \citep{webb_etal17}, therefore $\lambdam$ serves as an upper limit of the tidal strength. Alternatively, we discuss below also a case with
$$ \Omega_{\rm tid}^2 \approx \frac{1}{3} {\rm max}(\lambda_1, 0) $$
that ignores the effect of fully compressive tides with $\lambda_1<0$.

In the coordinate frame rotating with the cluster there is an additional contribution of the Coriolis force to the equipotential surface defining the tidal radius \citep{renaud_etal11}. This correction due to orbital rotation $\Omega_{\rm rot}^2$ is negligible for majority of our clusters. We analysed the rotational speed of cluster particles in our high-redshift discs and found that, due to the strong stellar feedback and hierarchical galaxy merging, the disc has very irregular shape and is dominated by turbulent motions. The typical rotational velocity is around $V_{\rm rot}\sim 50$~km/s, which corresponds to $\Omega^2\sim 100$~Gyr$^{-2}$ given the average disc size around 5~kpc. This term is in general smaller than the largest eigenvalue of the tidal tensor during most of the cluster lifetime, and therefore we ignore this term in the tidal calculation.

For brevity, we express the values of the tidal tensor and orbital frequencies in units of Gyr$^{-1}$. In some papers they are expressed in units of km~s$^{-1}$~kpc$^{-1}$, which are essentially interchangeable and differ only by a factor 1.02.

Once the strength of the tidal field, $\Omega_{\rm tid}$, is evaluated from the above approach, the decrease of the bound cluster mass is determined by \autoref{eq:bound-mass}. At $n$-th global time-step $dt_n$ in the simulations, we update the bound fraction as
$$ f_{\rm dyn}^{n+1} = \exp{\left(-dt_n/t_{\rm tid}\right)} \, f_{\rm dyn}^{n}. $$

\begin{figure}
\includegraphics[width=\hsize]{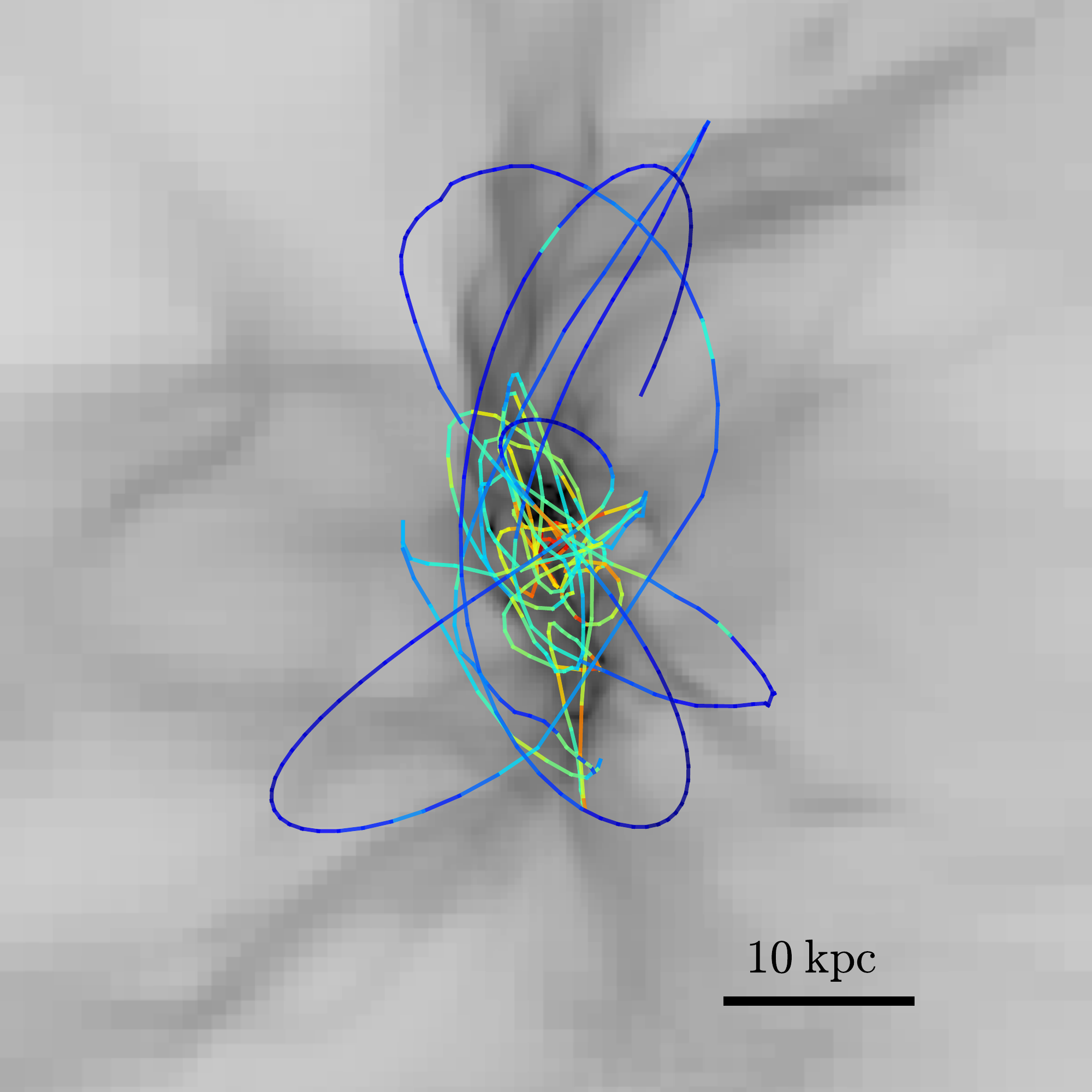}
\includegraphics[width=\hsize]{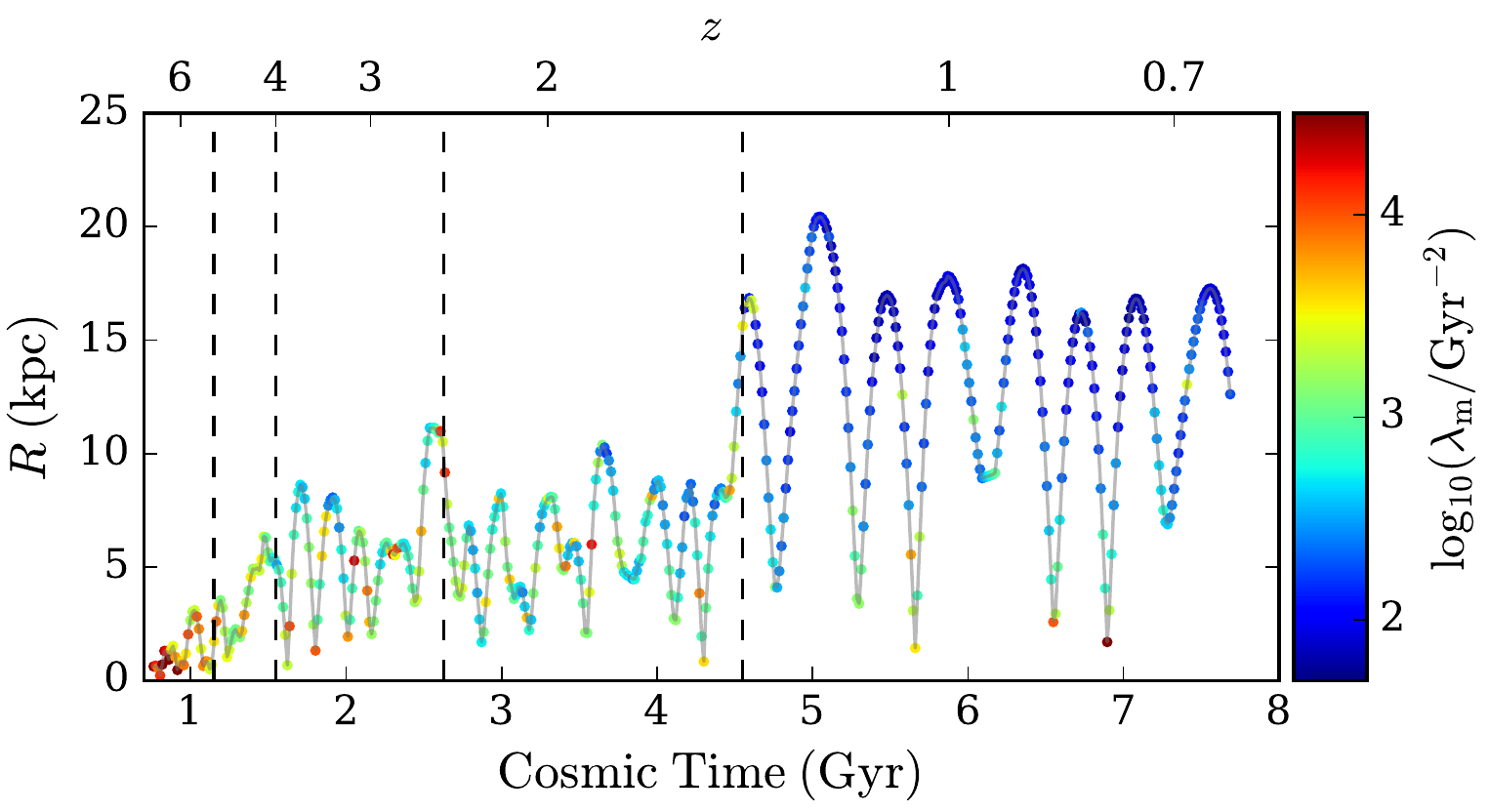}
\vspace{-4mm}
\caption{\small \textbf{Upper}: Orbit of a surviving cluster around the center-of-mass of the main galaxy in SFE200 run from $z=7$ to 0.6. Trajectory is coloured by the value of $\lambdam$ from the bottom panel. Gas surface density projected along the $z$-axis of the simulation box at $z\approx 0.6$ is shown as the background. Some glitches in the trajectory are due to sudden changes of the center-of-mass position of the host galaxy during mergers. \textbf{Lower}: Evolution of the galactocentric radius of the cluster shown in the upper panel. In both panels points are colour-coded by maximum absolute value of the eigenvalues of the tidal tensor, $\lambdam$. Epochs of four major merger events are labeled as vertical dashed lines.}\label{fig:orbit}
\end{figure}

\subsection{A typical cluster orbit around the main galaxy} \label{sec:orbits}

Upper panel of \autoref{fig:orbit} shows orbital motion of an example cluster that survives to the end of the SFE200 run. The selected cluster forms in-situ near the centre of the main progenitor but over time it migrates to higher orbit due to the interactions with dense structures in the disc and major merger events. The lower panel of \autoref{fig:orbit} quantifies the evolution of the galactocentric radius of this cluster until the last available simulation output at $z\approx 0.6$ (this run has been continued to lower redshift than the others, which stopped at $z\approx 1.5$). Initially, the cluster follows a roughly circular orbit until $z\approx 4.0$ when the first major merger occurs. From the lower panel, we find that the four major merger events at $z\approx 5.1$, 4.0, 2.5, and 1.4 significantly increase the orbital eccentricity and lift the cluster to a higher energy orbit. After the most recent merger at $z\approx 1.4$, the apocenter distance increases to 20~kpc, while the average pericenter distance remains within 3~kpc. Each time the cluster reaches its pericenter, it penetrates the gaseous disc of its host galaxy and experiences strong tidal forces. As we show in the next section, such a highly eccentric orbit leaves clear marks on the evolution of the tidal field experienced by the cluster.

\subsection{Tidal field evolution over cosmic time}\label{sec:results-tidal}

\begin{figure}
\includegraphics[width=\hsize]{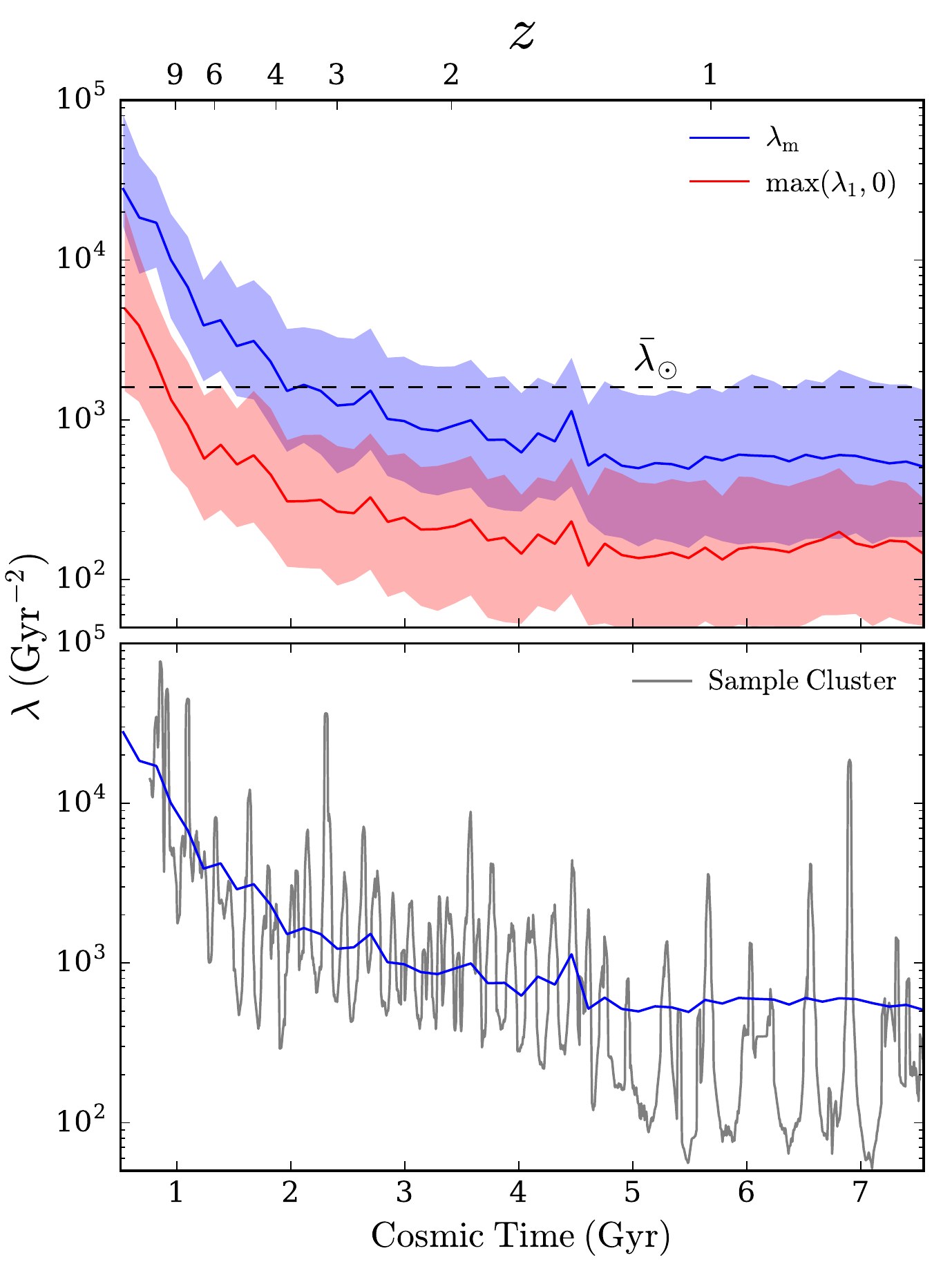}
\vspace{-4mm}
\caption{\small \textbf{Upper}: Time evolution of the eigenvalues of the tidal tensor, $\lambdam$ and $\lambda_1$, for clusters more massive than $10^5\Msun$ in SFE200 run. Blue and red lines show the median values of $\lambdam$ and of positive values of $\lambda_1$, respectively, as a function of cluster age, while the shaded areas show the 25-75\% interquartile range. Tidal strength at the orbit of the Sun, $\bar{\lambda}_{\odot}\approx1600\,\rm Gyr^{-2}$, is shown as dashed line for reference. \textbf{Lower}: Evolution of $\lambdam$ with high time resolution for the same sample cluster from \autoref{fig:orbit} (black line). Blue line reproduces the median trend from the upper panel.
}\label{fig:lambda}
\end{figure}

We show in \autoref{fig:lambda} the overall evolution of the tidal intensity around all clusters with mass above $10^5\Msun$. The tidal field is strongest immediately after clusters emerge from the natal cloud and orbit within the dense gaseous disc. The average amplitude of $\lambdam$ can reach very high values, $\sim10^5$~Gyr$^{-2}$ at high redshift, much higher than in the solar neighbourhood, which is typically estimated as $\lambda_\odot \equiv 2\, \Omega_\odot^2 \approx 1600$~Gyr$^{-2}$ \citep{renaud_etal17}. There is a clear trend that the tidal field weakens over time. The average value of $\lambdam$ decreases by more than an order of magnitude after the first two Gyr. Weakening of the tidal field reflects the orbital migration of the clusters. As we showed in previous section, star clusters that are formed in dense GMCs within the galactic disc are gradually scattered to the less dense outskirts of the galaxy. As clusters migrate to larger distance, the tidal field weakens.

Moreover, as expected, the median values of ${\rm max}(\lambda_1,0)$ are systematically lower than $\lambdam$ over the whole cosmic history. This is because $\lambdam$ can take the absolute value of the most negative eigenvalue of the tidal tensor $|\lambda_3|$, which is sometimes larger than $\lambda_1$. This is especially important for fully compressive tides when all eigenvalues are negative. We note that the type of tides alternates between compressive and extensive during cluster lifetime. We analyse the tidal evolution history of all cluster particles more massive than $10^5\Msun$. We find that, statistically, clusters experience compressive tides for roughly 31\% of their total lifetime. In the early stages when the cluster resides within the disc, tidal interactions are dominated by compressive tides. As the cluster migrates outward, the tides become more extensive. However, compressive tides occasionally reappear at lower redshift when the cluster penetrates the gaseous disc. We will show below that choosing $\lambdam$ or $\lambda_1$ for the tidal disruption calculation can lead to visible (though not critical) differences in the mass function and metallicity distribution of surviving clusters.

In the lower panel of \autoref{fig:lambda}, we show $\lambdam$ for the same cluster as in \autoref{fig:orbit} with higher time resolution. We find quasi-periodic oscillations of $\lambdam$ with prominent spikes repeated over a period of $\sim 100$~Myr, especially at later times. These oscillations are caused by the orbital motion around the galaxy, as can be seen in \autoref{fig:orbit}. Because of the highly eccentric orbit, each time the cluster approaches its pericenter, the cluster penetrates the dense gas disc and experiences a tidal shock that raises the eigenvalue by more than an order of magnitude.

We also examine the time evolution of the tides around model clusters in SFE50 and SFE100 runs and find that the trend and normalization is in general similar to that of the SFE200 run. This is because the global properties of the host galaxy for the three runs are very similar, which can be recognized by the star formation history and Kennicutt-Schmidt relation shown in Figure~5 and 6 of Paper II.

\begin{figure}
\includegraphics[width=\hsize]{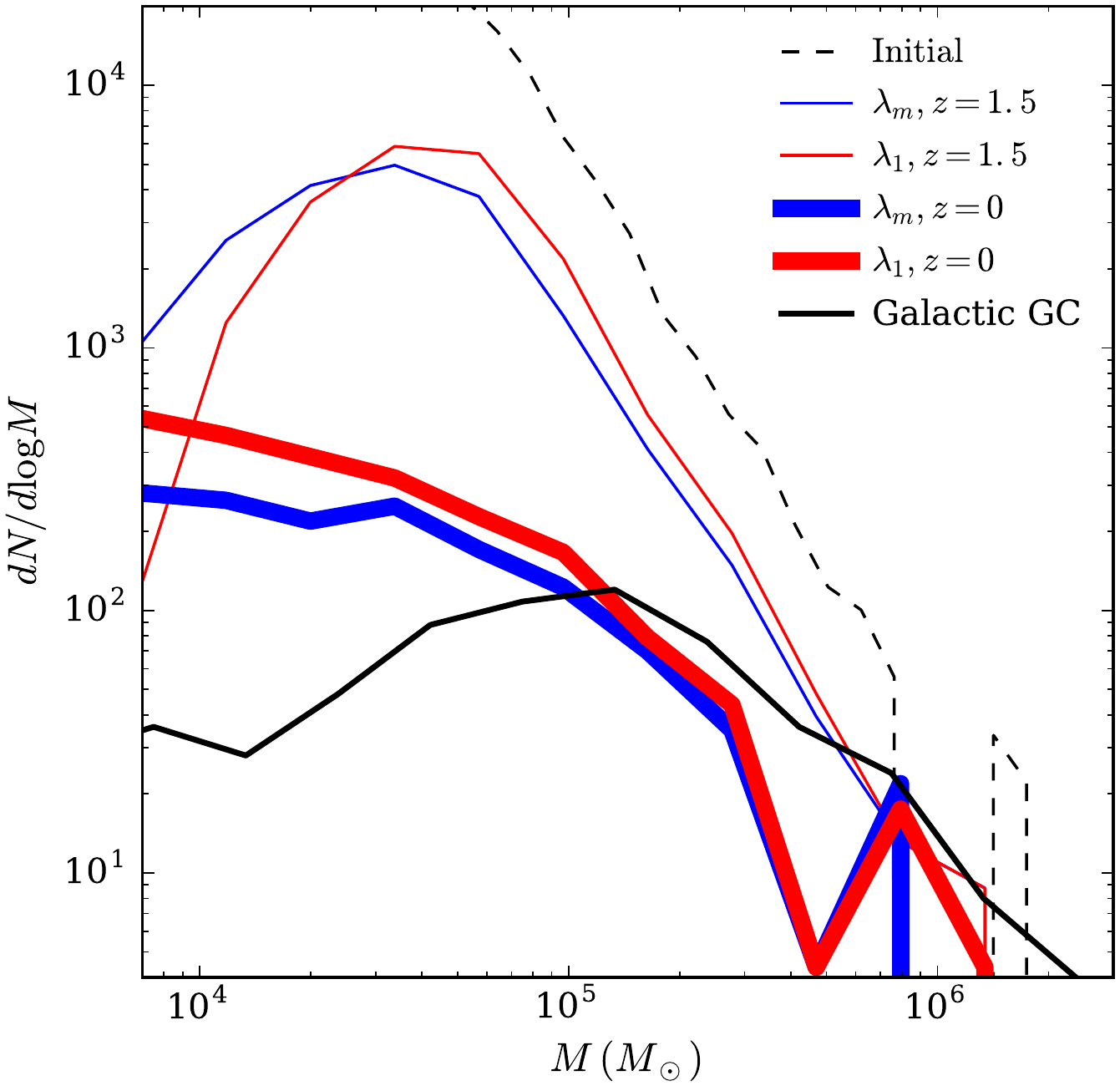}
\vspace{-4mm}
\caption{\small Mass function of surviving clusters at $z\approx1.5$ (thin lines) and $z=0$ (thick lines) for SFE100 run. colours show two ways of estimating the strength of tidal disruption: $\lambdam$ (blue) and $\lambda_1$ (red). The mass-loss from $z=1.5$ to $z=0$ is estimated with $\rm \Omega_{tid}=50~Gyr^{-1}$. The cluster initial mass function (CIMF) is shown as dashed line, which is truncated at low mass end due to the upper limit of the y-axis. Its full shape was shown in Figure~9 of Paper II. For reference, the observed mass function of the Galactic GCs is plotted as black line.
}\label{fig:MF_correction_tid50}
\end{figure}

\subsection{Evolution of the cluster mass function} \label{sec:results-massfunction}

The strong tidal field found above, especially soon after cluster formation, suggests that mass-loss via tidal disruption is critical to the evolution and survival of model clusters. 
Considering mass-loss due to stellar evolution and tidal disruption characterised by the eigenvalues of the tidal tensors, we now examine how the shape of the cluster mass function changes over time.

\autoref{fig:MF_correction_tid50} shows the distribution of several mass variables in run SFE100: initial bound cluster mass, bound mass at the last available simulation snapshot $M_{\rm bound}(z\approx1.5)$, and the projected bound mass at $z=0$. The latter is estimated by considering tidal disruption with a fixed value of $\Omega_{\rm tid}$ from $z=1.5$ to the present, as we do not have information on the actual strength of the tidal field in this interval. By the time of last output most clusters have migrated out of the galactic disc into the weak tides environment, which justifies this assumption. We adopt $\rm \Omega_{tid}=50~Gyr^{-1}$, which corresponds to the top $\sim5\%$ of the distribution of $\lambdam$ after $z=2$ in \autoref{fig:lambda}. Later we also consider a different value of $\Omega_{\rm tid}$ for comparison.

Tidal disruption at $z>1.5$ is estimated in two ways, based on either $\lambda_1$ or $\lambdam$. Since $\lambdam$ is the maximum absolute eigenvalue of the tidal tensor, it combines the contributions from both extensive and compressive tides and can be considered as an upper limit on the tidal disruption. On the other hand, the $\lambda_1$ case ignores the contribution of compressive tides and results in a lower limit. We find that the cluster mass function in the $\lambdam$ case is a suppressed compared to the $\lambda_1$ case, but the difference is relatively small.

The shape of the mass function changes significantly from an initial (truncated) power-law to a peaked, roughly log-normal function at $z=1.5$. Tidal disruption dramatically reduces the number of clusters of all mass, especially less massive than $\sim 10^5\Msun$. The projected mass functions at $z=0$ show further decrease in the number of massive clusters and increase at $M\lesssim 10^4\Msun$. Some clusters are not completely disrupted but simply lose a fraction of their mass and shift to the left of the distribution. 

To compare the mass function of surviving clusters with observations, we calculate the masses of Galactic GCs from the absolute V-band magnitude $M_V$ in \citet{harris96} with the luminosity-dependent mass-to-light ratio 
$$ \frac{M}{L_V} = 1.3 + \frac{4.5}{1+e^{2M_V+21.4}} $$
suggested by \citet{harris_etal17a}\footnote{Recenly, \citet{baumgardt17} suggested that most of the Galactic GCs have the mass-to-light ratios larger than 1.8. Adopting these ratios will potentially increase the discrepancy between the observed and model mass function.}, in their Eq.~(6). Relative to the observed mass function of Galactic GCs, our model underpredicts the number of massive clusters and overpredicts the number of clusters with $M<10^5\Msun$. Since the simulations do not run past $z=1.5$, they do not include any massive clusters that could form at later epochs. This could partially compensate the difference with the observations for high-mass clusters. For the low-mass clusters, the model result is in excess of the observations and does not reproduce the peak of mass function at $M\sim10^5\Msun$. This discrepancy could be due to a number factors: assumption of constant $\Omega_{\rm tid}$ at late epochs; not including cluster sizes in the disruption calculation; or a lack of detailed treatment of tidal shocks. Interestingly, we also find that low-mass clusters in the simulations usually form in low-density regions with lower $\Omega_{\rm tid}$, and thus preferentially experience less initial tidal disruption than the high-mass clusters.

Note that our previous semi-analytical models of GC formation \citep{muratov_gnedin10, li_gnedin14, choksi_etal18} reproduce the observed log-normal mass function using a different disruption time-scale that depends linearly on mass, $t_{\rm tid} \propto M$. However, there are several differences between these semi-analytical models and the current simulations. The semi-analytical models adopt a power-law CIMF with an index of -2, while the current simulations obtain more realistic CIMFs that depend on the local gas properties within the galactic disc (Papers I and II). Moreover, since the semi-analytical model does not contain orbital information for model clusters, it has to assume a time-independent $\Omega_{\rm tid}$ around 40-100~Gyr$^{-1}$. In the simulations, the local tidal field for each cluster is calculated in the run-time. The sub-linear slope of the mass dependence adopted in this paper, $t_{\rm tid}\propto M^{2/3}$, slows down the disruption of low-mass clusters at the same $\Omega_{\rm tid}$ and causes the excess number of clusters at low-mass end.

\begin{figure}
\includegraphics[width=\hsize]{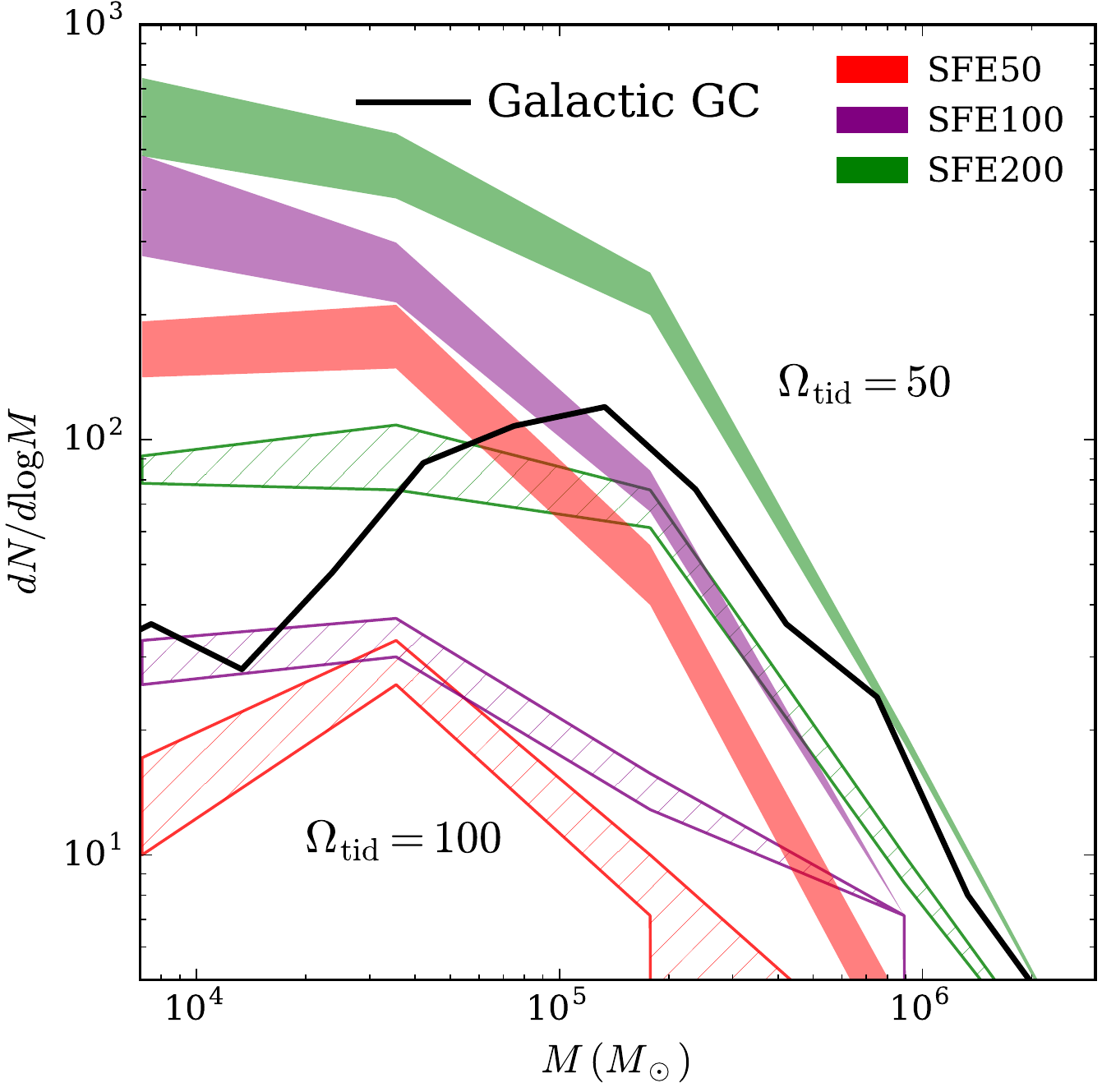}
\vspace{-4mm}
\caption{\small Mass function of surviving clusters at $z=0$ for runs SFE50, SFE100, SFE200 using $\rm \Omega_{tid}=50\; or\; 100~Gyr^{-1}$ in solid and line-shaded contours, respectively. Contours for each run represent the range of the mass function using two ways of estimating the strength of the tidal field ($\lambdam$ and $\lambda_1$). No cluster survived in SFE10 and SFEturb runs, due to their small initial bound fractions. Mass function of the Galactic GCs is plotted as black line.
}\label{fig:MF_z0}
\end{figure}

\autoref{fig:MF_z0} compares the predicted mass functions at $z=0$ in three runs: SFE50, SFE100, and SFE200. They have similar shapes but different normalizations, because of the different initial bound fractions of clusters. We find that the normalization of the model mass function correlates strongly with the choice of $\epsff$, which controls the normalization of $f_{\rm i}$. For example, in SFE200 run, the number of surviving clusters is three times larger than the number of Galactic GCs, while in SFE10 run no cluster survives until $z=0$ at all. The SFE100 run has the closest match to the observed mass function, although as discussed above, it overpredicts the number of less massive ($<10^5\Msun$) GCs and underpredicts more massive ($>10^5\Msun$) ones. This conclusion relies directly on the sub-grid model of the initial bound fraction.

The mass function of surviving clusters at $z=0$ depends also on the assumed tidal strength after the last simulation output. In \autoref{fig:MF_z0} we experiment with two values of $\rm \Omega_{tid}=50\;or\;100\,Gyr^{-1}$. The motivation for these specific choices is empirical -- smaller values would lead to even less disruption of low-mass clusters. Therefore, the effect of smaller values of $\Omega_{\rm tid}$ can be extrapolated from the results presented in \autoref{fig:MF_z0}. In the case of $\rm \Omega_{tid}=100\,Gyr^{-1}$, the number of surviving clusters drops dramatically and all available runs underpredict the number of observed Galactic GCs. We therefore set $\rm \Omega_{tid}=50\,Gyr^{-1}$ as a fiducial value for the rest of the paper. \autoref{tab:GC} lists the total number and mass of GC candidates and surviving GCs in three runs for $\rm \Omega_{tid}=50\,Gyr^{-1}$. The values are obtained from the tidal disruption calculations using $\lambda_1$. The GC candidates are defined as clusters that have initial mass larger than the lowest initial cluster mass that managed to survive to $z=0$ ($\sim8\times10^4\Msun$), while the surviving GCs are defined as clusters with residual mass above $10^4\Msun$ at $z=0$. We find that roughly 10\% of total mass of GC candidates survives to the present day. The total mass of surviving clusters at $z=0$ ranges from $1.5\times10^7\Msun$ to $6\times10^7\Msun$. Run SFE100 is most consistent with the observed GC system mass for Milky Way-sized galaxies \citep[e.g.][]{spitler_forbes09}.

\begin{table*}
    \centering
    \begin{tabular}{lccc}
    \hline
        Run & GC candidates & Surviving GCs & Number/Mass fraction of surviving GCs\\
    \hline
        SFE50 & 708 / $1.23\times10^8\Msun$ & 250 / $1.48\times10^7\Msun$ & 0.35 / 0.12 \\
        SFE100 & 1688 / $2.53\times10^8\Msun$ & 369 / $2.49\times10^7\Msun$ & 0.22 / 0.10 \\
        SFE200 & 2450 / $4.36\times10^8\Msun$ & 733 / $5.99\times10^7\Msun$ & 0.30 / 0.14 \\
    \hline
    \end{tabular}
    \caption{Total number and mass of GC candidates and surviving GCs in three runs for $\rm \Omega_{tid}=50\,Gyr^{-1}$.}
    \label{tab:GC}
\end{table*}

\section{Metallicity Distribution}\label{sec:metal}

One of the most important property of star clusters is their metallicity, because it does not change as the cluster evolves. Knowing the distribution of clusters expected to survive to the present allows us to derive their observable metallicity distribution. In this section we describe how we calculate the metallicity of the model clusters.

\subsection{Chemical enrichment in the simulations} \label{sec:metallicity}

In our simulations, young clusters launch feedback by depositing mass, momentum, energy, and metals to the ambient medium. We record metal enrichment using two separate variables, $M^{ZII}$ and $M^{ZIa}$, that account for the contributions of supernovae Type II and Type Ia, respectively. When a cluster particle accretes mass $m_i$ at a given time-step $i$, a fraction of this mass $m_i^Z = Z_i\,m_i$ is contributed by the metals, based on the metallicity $Z_i$ of the accreted gas, separately for each SN type. The mass of metals for a given cluster particle increases from step $i-1$ to step $i$ as $M_i^Z = M_{i-1}^Z + m_i^Z$. 

The metallicity of specific elements is calculated by post-processing using metal yield tables from \citet{nomoto_etal06} and \citet{iwamoto_etal99} for SNII and SNIa, respectively. For example, when converting the total to iron metallicity of star particles in the simulations, we find that the values of the offset $\feh-\ZH$ have a typical range from $-0.18$ to 0.0~dex.

Compared to Paper I, in new runs presented in Paper II and here we reduced the maximum stellar mass that triggers SN Type II from $100\Msun$ to $40\Msun$, since main sequence stars more massive than roughly $40\Msun$ cannot launch successful SNe and distribute metals to the ambient medium \citep[e.g.][]{heger_etal03}. The new choice of the maximum mass is also consistent with the value used in previous cosmological simulations \citep[e.g.][]{agertz_etal13}. This change has little effect on the strength of energy and momentum feedback since the number of SNe is dominated by the lower-mass stars. However, the total mass of SN ejecta, including the metals, is dominated by the highest-mass stars. For a single stellar population with \citet{kroupa01} IMF, this change of the maximum stellar mass reduces the SNII yield by $0.23$~dex.

\subsection{Self-enrichment} \label{sec:metallicity-self}

\begin{figure*}
\includegraphics[width=1.0\hsize]{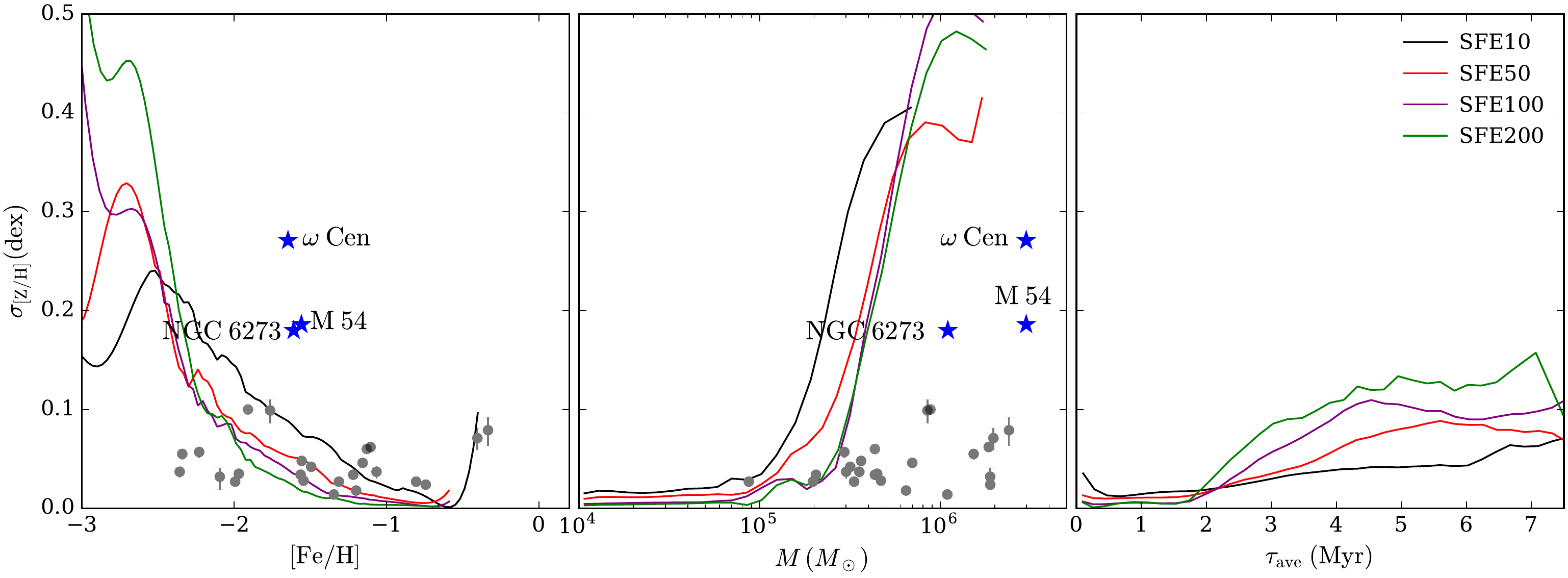}
\vspace{-4mm}
\caption{\small Intrinsic spread of $\ZH$, estimated by Eq.~\ref{eq:metal-spread}, as a function of cluster metallicity (left), initial cluster particle mass (middle), and cluster formation time-scale (right). Solid lines show the median value of $\sigma_\ZH$ in bins. colours represent different runs as described in the legend. Symbols with errorbars show the observed iron spread for Galactic GCs from several different observations from \citet{carretta_etal09, da_costa_etal14, johnsonC_etal17a, johnsonC_etal17b}. Three GGCs with highest iron spread, $\omega$-Cen, M54, and NGC 6273, are highlighted by blue stars. We assume the logarithmic iron spread to be the same as the spread of total metallicity. Initial cluster masses are estimated by \citet{balbinot_gieles18} by correcting the observed masses for the dynamical evolution.
}\label{fig:metal-dispersion}
\end{figure*}

Although GCs have traditionally been thought to have single stellar populations with the same age and metallicity, recent spectroscopic observations have accumulated evidence of anomalous abundances of light elements in most Galactic GCs \citep[see a recent review by][]{bastian_lardo18}. These may be interpreted as multiple populations of stars, although for iron the abundance spread is consistent with observational errors ($\lesssim 0.05\,$dex).
In addition, a few most massive clusters show a significant spread of iron abundance, such as $\omega$~Centauri \citep[e.g.][]{johnson_pilachowshi10} and M54 \citep{carretta_etal10}.
While it is generally believed that the clusters with large iron spread is the remnant nucleus of a disrupted dwarf galaxy \citep[e.g.][]{da_costa16}, the physical origin of the multiple populations for other GCs is still highly debatable.

Are these clusters formed from a single episode of star formation? Could self-enrichment explain the observed metallicity spread for some GCs?
We can investigate the possible origin of the metal spread due to the continuous gas accretion and self-enrichment during the course of cluster formation by calculating the metallicity spread of cluster particles in our simulations. We introduce an additional variable for each cluster, to record the spread of metal abundance from Type II SNe \footnote{Type Ia SNe are irrelevant here, since the delay time between the formation of stars and SN explosion is typically on the order of Gyr, which is significantly longer than the formation time-scale of star clusters.} during the active phase of cluster formation.

The variance of a random variable $X$ is defined as 
$$ \sigma_X \equiv \frac{1}{N} \sum_{i=1}^N(x_i-\bar{x})^2, {\rm \;where\;} \bar{x}\equiv \frac{1}{N} \sum_{i=1}^N x_i. $$
For a more complicated case with weights $w_i$ associated with each point $x_i$, the weighted variance can be derived as 
$$ \sigma_X = \frac{\sum_{i=1}^N w_i(x_i-\bar{x})^2}{\sum_{i=1}^N w_i}, {\rm \;where\;} \bar{x}\equiv \frac{\sum_{i=1}^N w_i x_i}{\sum_{i=1}^N w_i} $$
is the weighed mean. For a straightforward way to calculate the variance of metallicity $Z=M^Z/M$ during cluster formation, the accreted mass part $m_i$ and its associated metallicity $Z_i=m_i^Z/m_i$ at each local time-step need to be stored, which is computationally prohibitive. Instead, we adopt another way of recording the mass-weighted metallicity variance without using the full accretion history. We introduce an additional variable, $M^{ZZ} \equiv \sum m_i Z_i^2$, attached to each cluster particle. The metallicity variance can then be expressed as
\begin{align*}
  \sigma_Z^2 &= \frac{\sum m_i (Z_i-\bar{Z})^2}{\sum m_i} \\
  &= \frac{\sum m_i Z_i^2}{\sum m_i}-\left(\frac{\sum m_iZ_i}{\sum m_i}\right)^2 \\
  &= \frac{M^{ZZ}}{M} - \left(\frac{M^Z}{M}\right)^2.
\end{align*}
Therefore, instead of recording the whole mass and metal accretion histories, the variance of $Z$ can be fully recovered by using $M$, $M^Z$, and $M^{ZZ}$.

In observations, metallicity is usually quoted as the logarithm of the ratio between the abundance of stars and the solar abundance, $\ZH$. The variance of $\ZH$ is therefore different from the variance of $Z$. Fortunately, we have found an analytical conversion between the two definitions so that no additional variables need to be stored during simulation runtime to recover both variances. We analysed the accretion history of mass and metals for many model clusters and found that the abundance of the accreted material follows a log-normal distribution. For a random variable $X$ following a log-normal distribution with mean $\mu$ and variance $\sigma^2$, the mean and variance of $X$ are 
$$
  \mu_X = e^{\mu+\sigma^2/2}, \quad
  \sigma_X = (e^{\sigma^2}-1) \; e^{2\mu+\sigma^2}.
$$
Therefore, for given $\mu_X$ and $\sigma_X$, the variance of $\ln{X}$ can be calculated as $\sigma^2=\ln{(1+\sigma_X^2/\mu_X^2)}$. Applying this relation to our particular case where $\ZH = \rm \log_{10}(Z/Z_{\sun})$, we have
\begin{equation}
  \sigma^2_{\rm [Z/H]}=\ln{(1+\sigma_{Z}^2/\mu_Z^2)}\, /(\ln{10})^2.
  \label{eq:metal-spread}
\end{equation}

\subsection{Metallicity spread via self-enrichment} \label{sec:results-metaldispersion}

Our simulations resolve the formation of each cluster with thousands of time-steps, and therefore, can trace the metal accretion history during its growth. Using the method described in the last section, for each cluster particle we record the internal metal spread in linear ($\sigma_Z$) and logarithmic ($\sigma_\ZH$) space. \autoref{fig:metal-dispersion} shows correlations between the median metal spread in log space and cluster metallicity $\ZH$, initial cluster particle mass, and cluster formation time-scale $\tau_{\rm ave}$, for all clusters formed in four runs. The typical 25-75\% interquartile range for $\sigma_\ZH$ is around 0.03~dex. The range becomes larger ($\sim0.1$~dex) for the most metal-poor ($\ZH<-2.5$) and most massive ($M>2\times10^5\Msun$) clusters, due to small number of clusters in each bin.

In general, we find small metal spread $\sigma_\ZH<0.1$ for clusters with metallicity $\ZH>-2.0$, in rough agreement with the observations of Galactic GCs. The only exceptions are three Galactic GCs, $\omega$~Cen, M~54, and NGC~6273, which have iron spread $\gtrapprox0.2$~dex. These three GCs, however, are suspected to be the debris of disrupted dwarf galaxies. We also find a clear trend that metal spread increases with decreasing cluster metallicity. Clusters with $\ZH=-2.5$ have a median spread $\sigma_\ZH\approx 0.2-0.4$~dex, while $\ZH>-1.0$ clusters show negligible spread. Since the spread is caused by the accretion of gas with different metallicity, the reason for this correlation lies in the definition of logarithmic metallicity. It requires more metals to contaminate metal-rich clusters by a given amount in dex as for the metal-poor ones, and is therefore harder. Such a trend, however, is not obvious in the observed GCs. One possible explanation for the discrepancy is that the available observations are biased toward GCs with $\ZH>-2.0$ -- their metallicity spread is so small that the trend cannot be seen easily. Accurate metal spread measurements of $\ZH<-2.0$ GCs are needed to test our model. A caveat that needs to be emphasized is that the metal spread recorded in the simulations comes from SNe II, which are less efficient at producing iron than SNe Ia. Comparing the spread of total metallicity in the simulations to the observed iron spread may not be reliable.

We explore the dependence of the metallicity spread on cluster mass in the middle panel of \autoref{fig:metal-dispersion}. Massive clusters ($>10^5\Msun$) are preferentially formed at high redshifts when the host galaxy experiences more frequent major mergers. These clusters are also typically metal-poor. Thus the anti-correlation of the metal spread with metallicity translates to a strong positive correlation with cluster mass, especially at $M \gtrsim 3\times 10^5\Msun$. In a parallel study, \citet{brown_etal18} investigated the metal spread in nuclear star clusters formed at the centers of galaxies in our simulation suite. These systems are composed of several stellar particles within a region of 10-20 pc in radius. Their metallicity distribution is broadened by the internal spread within each particle and the external spread among the particles. Even these composite systems have typical spreads of 0.1~dex, with the largest reaching 0.4~dex.

In the right panel of \autoref{fig:metal-dispersion}, we also find that the metal spread increases with cluster formation time-scale. The latter is defined in Paper I as
$$\tau_{\rm ave} \equiv \frac{\int t\dot{M}(t)dt}{\int \dot{M}(t)dt},$$
where $\dot{M}(t)$ is the mass growth rate of a cluster at time $t$. This $\tau_{\rm ave}$ can be considered a characteristic time-scale to grow the cluster to its final mass. Clusters with longer $\tau_{\rm ave}$ are more likely to be exposed to a variety of enriched gas parcels and thus acquire larger metal spread.

\subsection{Stellar mass - metallicity relation of host galaxy} \label{sec:results-massmetal}

\begin{figure}
\includegraphics[width=1.0\hsize]{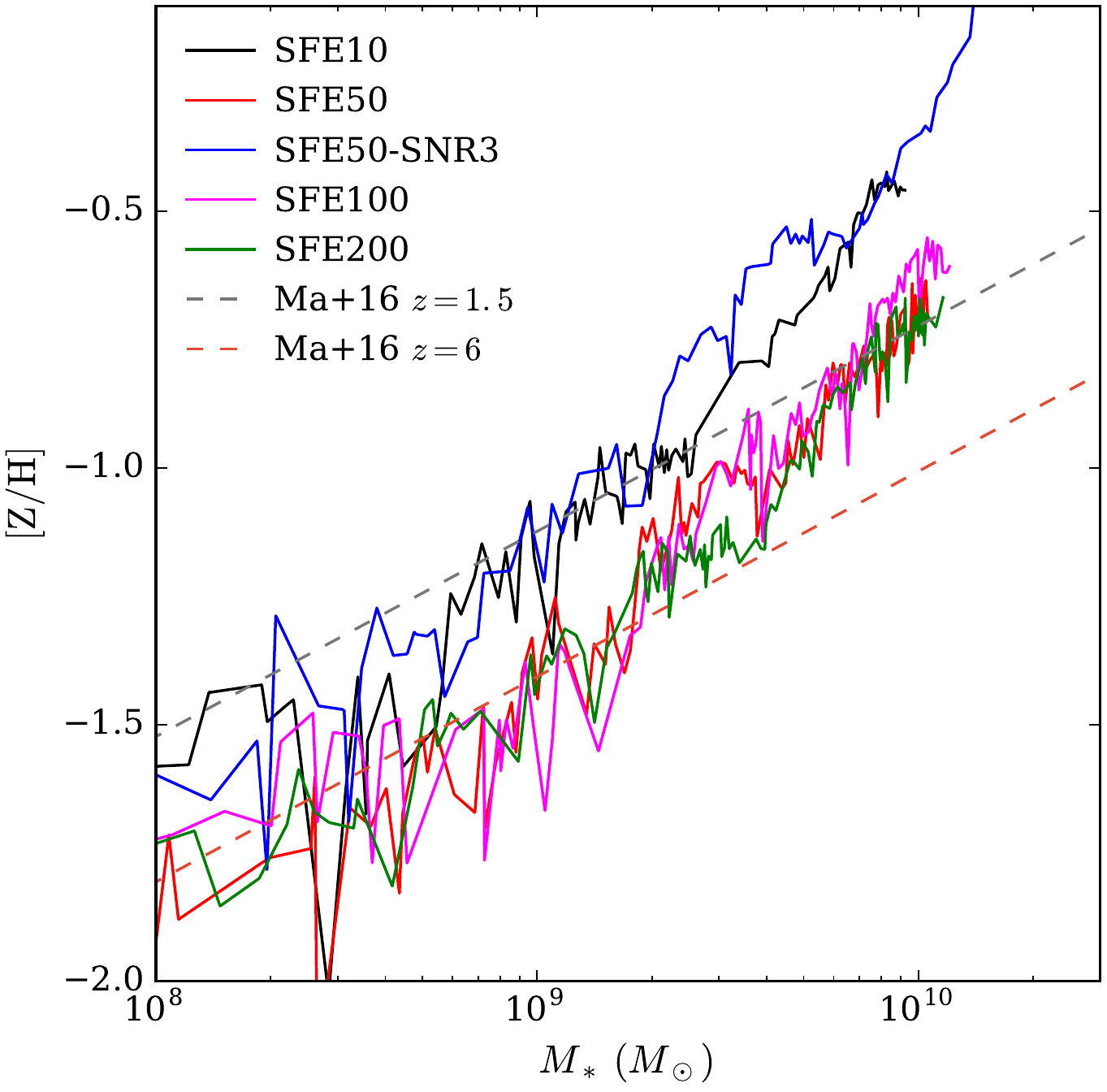}
\vspace{-4mm}
\caption{\small Evolution of stellar mass-stellar metallicity relation of the main galaxy for runs with different star formation and feedback parameters. Each line represents a time sequence of one galaxy, from $z\approx 6$ to the final available output at $z \gtrsim 1.5$. The stellar mass of the main galaxy is the total mass of cluster particles within $0.5 R_{\rm vir}$. For comparison, the stellar mass-stellar metallicity relation at $z=1.5$ and 6 from the FIRE simulation \citep{ma_etal16} is overplotted as dashed lines.
}\label{fig:metal_history}
\end{figure}

Before examining the metallicity distribution of surviving star clusters, we first investigate the growth of the average stellar metallicity of the main galaxy across cosmic time. \autoref{fig:metal_history} shows the stellar mass-metallicity relation (MMR) in different runs of our simulation suite. The stellar mass is obtained using all particles within half of the virial radius of the main galaxy (large enough radius to include all stars belonging to the central galaxy and closest satellites). Each line shows the evolutionary track of the main galaxy from $z\approx 6$ to $z\approx 1.5$. We find a strong positive correlation between galaxy metallicity and stellar mass, consistent with the observations of stellar metallicity in different types of galaxies \citep[e.g.][]{gallazzi_etal05, kirby_etal13}. Note that the trend shown here results from a combination of two effects: the galaxy moving up the MMR at a given epoch due to the growth of its mass, and the evolution of the normalization of the MMR with redshift.

It is challenging to obtain the stellar MMR from observations of galaxies at $z>1.5$. Instead we compare our result with that of the FIRE simulations presented in \citet{ma_etal16}, who predicted an exponential redshift evolution of the MMR at $z=0-6$: $\ZH=0.4\log{(\Ms/10^{10}\Msun)}+0.67\exp{(-0.5z)}-1.04$. In our runs with $\epsff\geqslant 0.5$ and default feedback boosting factor, the main galaxy with $\Ms\sim 10^{8}\Msun$ at $z=6$ has an average metallicity $\ZH\approx -1.7$. As the galaxy grows its mass, its average metallicity becomes larger and, at $z=1.5$, the main galaxy with $\Ms\sim 10^{10}\Msun$ has a total metallicity $\ZH\approx -0.7$. Both the normalization and the evolution of this relation are similar to that obtained in the FIRE simulations.

Run SFE50-SNR3 with weaker feedback produces metallicity systematically higher by 0.3~dex for $\Ms<10^{10}\Msun$. For galaxies of stellar mass above $10^{10}\Msun$, the metallicity rises to even super-solar values. This is caused by a strong starburst concentrated at the centre of the galaxy, which produces metals that cannot be efficiently dispersed to larger radii due to inefficient feedback. Run SFE10 also shows some metal enhancement over the runs with $\epsff\geqslant 0.5$, possibly because star formation is slow and cannot create sufficient gas outflows. In contrast, the three higher-efficiency runs have the MMR consistent with each other.

\begin{figure}
\includegraphics[width=1.0\hsize]{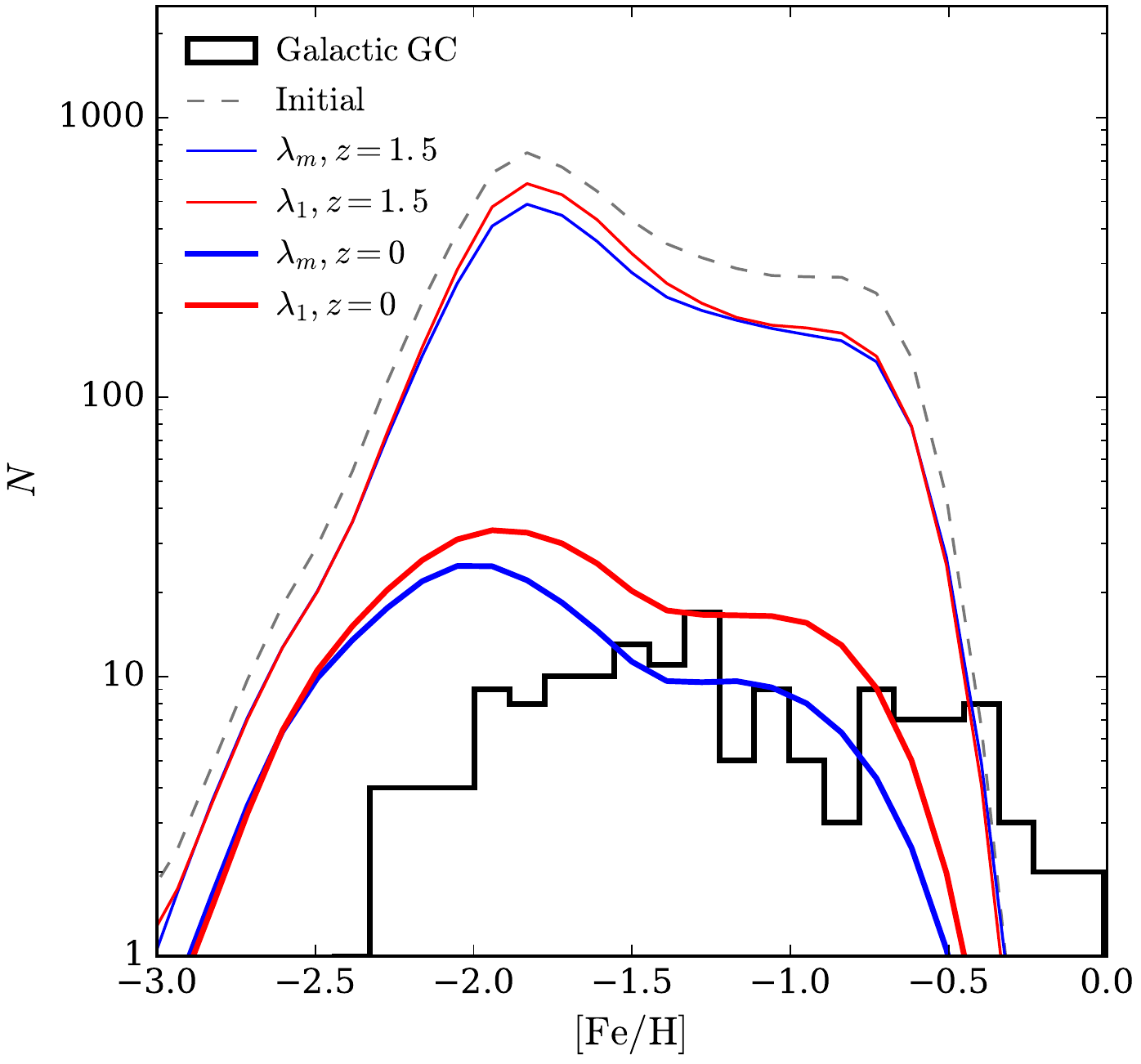}
\vspace{-4mm}
\caption{\small Metallicity distribution of star clusters in the main galaxy in run SFE100. Clusters with initial mass larger than $10^5\Msun$ are shown as black dashed line, while clusters that survive (with remaining mass $>10^4\Msun$) at $z=1.5$ and $z=0$ are shown as thin and thick solid lines, respectively. Final bound cluster mass is calculated via two methods: $\lambdam$ (blue) and $\lambda_1$ (red). Metallicity distribution of Galactic GCs is shown as black solid line for reference.
}\label{fig:metal_dist}
\end{figure}

\subsection{Metallicity distribution of surviving clusters} \label{sec:results-metal}

As shown in \autoref{sec:results-massfunction}, clusters less massive than $10^5\Msun$ in our simulations are fully disrupted; only about 10\% of GC candidate mass remains bound until the present in SFE100 run. Dynamical disruption modifies the metallicity distribution of bound clusters over time, but it is a smaller effect than for the mass function.

\autoref{fig:metal_dist} shows the metallicity distribution of clusters with initial mass above $10^5\Msun$ within the virial radius of main galaxy at $z=1.5$ for run SFE100. The distribution shows a broad range of metallicity from $\feh=-2.8$ to $-0.5$. Dynamical disruption destroys majority of these clusters: about $97\%$ of the metal-poor ones ($\feh<-1.6$) and $90\%$ of the metal-rich ones. Metal-poor clusters typically form at higher redshifts than the metal-rich ones, and therefore they have more time to experience tidal mass-loss.

Interestingly, at high redshift the host galaxy experiences more violent and frequent mergers, which dissipate angular momentum of the gas, create large-scale turbulent motions, and generate dense massive gas clumps, which in turn produce massive star clusters. This enhancement of massive cluster formation at high redshift compensates for the longer tidal mass-loss, resulting a roughly flat metallicity distribution of surviving clusters over the metallicity range $\feh=[-2.3,-0.7]$.

The metallicity distribution of surviving clusters is generally consistent with the observations of the Galactic GCs, although the modelled distribution systematically underestimates $\feh$ by $\sim0.2$~dex. The most metal-rich GCs are missing in the simulation. Given that the last simulation output reached only $z\approx1.5$, we may expect more metal-rich clusters to be formed during later epochs.

\begin{figure}
\includegraphics[width=\hsize]{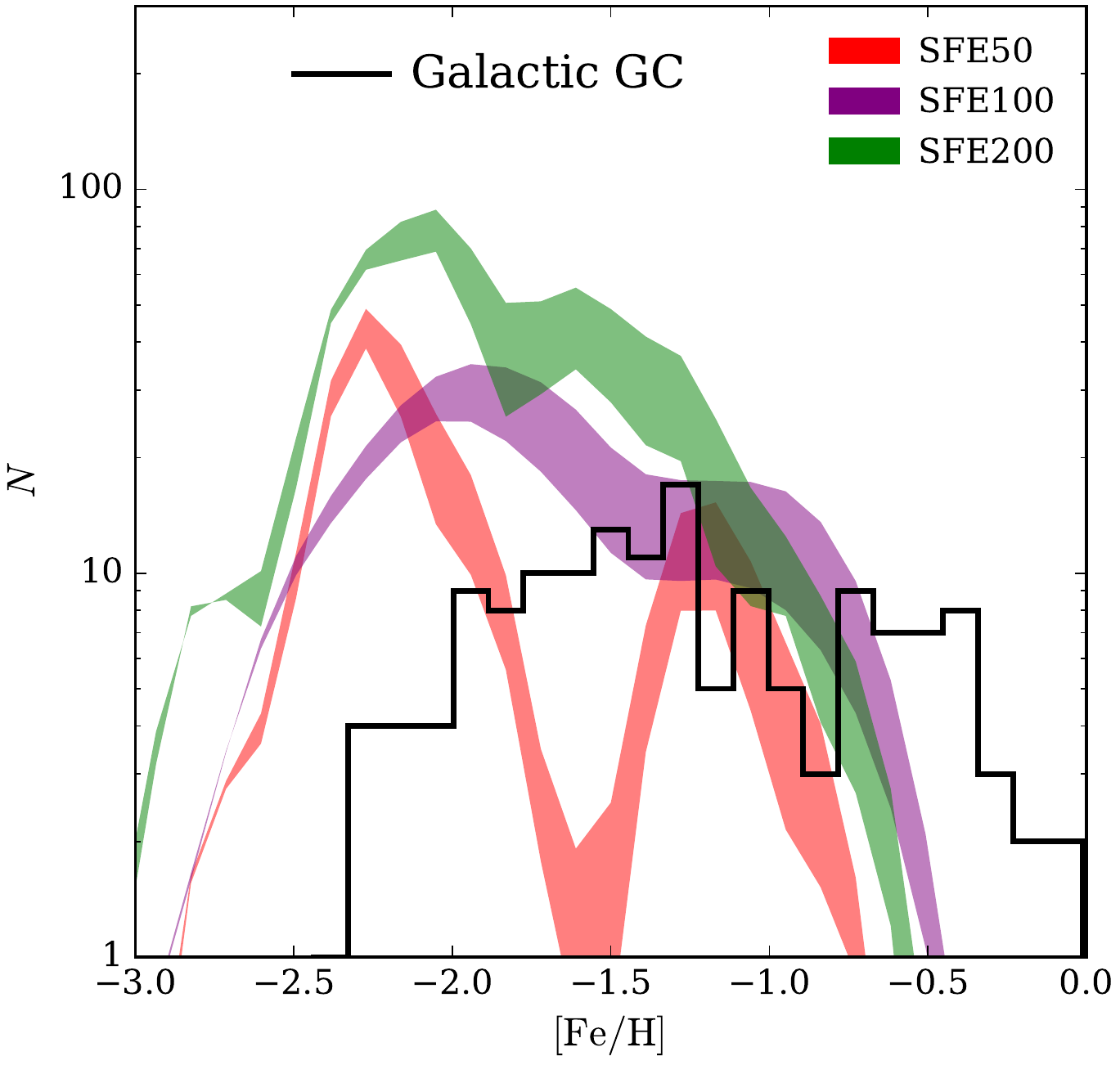}
\vspace{-4mm}
\caption{\small Metallicity distribution of surviving star clusters (with remaining mass $>10^4\Msun$) in the main galaxy at $z=0$ in runs SFE50 (red), SFE100 (purple), and SFE200 (green). Contours for each run envelop the distribution using two ways of estimating the strength of the tidal field ($\lambdam$ and $\lambda_1$). Metallicity distribution of Galactic GCs is shown as black line for reference.
}\label{fig:metal_z0}
\end{figure}

\autoref{fig:metal_z0} compares the metallicity distributions of surviving clusters for three runs. We again use the two different ways of calculating the tidal field at $z\gtrsim1.5$ (via $\lambda_1$ and $\lambdam$) to set the upper and lower contours. At lower redshifts we set the tidal strength to be $\rm \Omega_{tid}=50\,Gyr^{-1}$. As in \autoref{fig:MF_z0}, the normalization of the distribution increases with $\epsff$ because of the positive correlation between $\epsff$ and initial bound fraction $f_{\rm i}$. All three distributions cover roughly the same metallicity range because of the similarity of the galactic MMR in the three runs. All runs overpredict the number of metal-poor clusters and underpredict the number of metal-rich ones. The shape of these distributions is determined by the chemical enrichment and cluster formation physics, but not by the value of $\epsff$ assumed in the simulation.

\section{Discussion} \label{sec:discussion}

\subsection{Galactic mergers as triggers of formation of most massive star clusters}

We showed in Paper I that major mergers of galaxies enhance the rate of formation of most massive clusters. Here we can quantify this with our updated model. We find that majority (75\%) of clusters with $M>2\times10^5\Msun$ form within 200~Myr of the three major mergers, which happened at $z\approx 6.3$, 5.1, and 4.0. A large fraction of these massive clusters survive to the present and become GCs. In contrast, the formation of clusters with $M<10^5\Msun$ does not correlate with major-merger events but follows instead the overall star formation history of the host galaxy.

The merger-driven GC formation scenario has been considered for a long time \citep{ashman_zepf92} and used to build analytical models for the formation and evolution of GC populations in different types of galaxies \citep{muratov_gnedin10,tonini13,li_gnedin14,choksi_etal18,choksi_gnedin18}. Here we find additional support for this scenario in our cosmological hydrodynamic simulations. Moreover, as shown in \autoref{fig:orbit} and \autoref{sec:orbits}, major mergers also transport some fraction of clusters from the galactic disc to the halo, where the tidal field is significantly weaker. This merger-driven dynamical migration slows down GC disruption,
as has been suggested by \citet{kruijssen15}.

\subsection{Tidal evolution across cosmic time}

modelling tidal disruption in realistic galactic environment is crucial for predicting the mass and metallicity distributions of surviving clusters. Recent efforts have been made to track the tidal history along cosmic time in cosmological hydrodynamic simulations \citep[e.g.][]{renaud_etal17,pfeffer_etal18}. Here, we compare our results with previous studies, highlight some similarities and differences, and comment on possible reasons of the discrepancy.

Using similar cosmological initial conditions, \citet{renaud_etal17} presented a zoom-in simulation of a Milky Way-sized galaxy and studied the tidal evolution history of their chosen GC candidates. They found that the tidal strength stays roughly constant until $z\approx1.5$, when it begins to rise toward lower redshift. They argue that young massive clusters are less likely to be disrupted until late epochs when the clusters are assembled in a massive central galaxy. Note that their conclusions are based on a subset of stellar particles, since their simulation does not have explicit cluster formation physics, and therefore tidal histories so extracted represent the average tidal strength of the whole galaxy rather than that of the massive clusters. In reality, massive clusters are preferentially formed in dense environments and gradually migrate to higher orbits when they are scattered by dense clumps within the galactic disc as well as by major merger events.

Using simulations with a specifically designed cluster formation model, \citet{pfeffer_etal18} found gradual increase of tidal strength with increasing lookback time, similar to our result presented here. This general agreement confirms the importance of modelling of massive cluster formation when investigating their subsequent evolution. Quantitatively, however, both \citet{renaud_etal17} and \citet{pfeffer_etal18} obtained much lower tidal strength than that found in our simulations. For example, in \citet{pfeffer_etal18}, the median tidal strength is around $\lambdam\sim 100$~Gyr$^{-2}$, which is about two orders of magnitude smaller than our results during the first Gyr of cosmic time shown in \autoref{fig:lambda}. Even the maximum values of their $\lambdam$ are usually smaller than $10^4$~Gyr$^{-2}$, while the maximum values for our clusters can reach as high as $10^7$~Gyr$^{-2}$ for the first Gyr of cosmic time and stays around $10^5$~Gyr$^{-2}$ later on. The tidal strength around a cluster depends on the local density field resolved by the simulations. Their simulations are based on the EAGLE cosmological simulations, which have $\sim$kpc scale gravitational softening length. The maximum gas density that EAGLE can reach is around $10^3$~cm$^{-3}$ \citep{crain_etal17}, while in our simulations the gas density can be as high as $10^5-10^6$~cm$^{-3}$. The simulation of \citet{renaud_etal17} suffers from similar limitations, with a $\sim200$~pc spatial resolution to derive the tidal tensor around star particles.

The two orders of magnitude higher tidal strength in our simulations translates into one order of magnitude shorter disruption time-scale. Such a difference can lead to dramatic changes in the predicted survival rate of GCs and in the final mass and metallicity distributions. We emphasis again that, to fully resolve tidal disruption of GCs orbiting around their host galaxies, simulations need to have spatial resolution that is comparable to the GMC scale, from which young massive clusters emerge. Lower-resolution simulations would inevitably underestimate the tidal strength, especially at early times when the clusters reside within a dense gaseous disc.

\subsection{Initial bound fraction and model variations}

Both the mass and metallicity distributions of surviving clusters are affected by the choice of $\epsff$ in our simulations. This dependence is caused by the subgrid model for the initial bound fraction as a function of integrated star formation efficiency of individual clouds; see Paper II for more details. In the current model, we adopted a linear dependence of $f_{\rm i}$ on the local integrated star formation efficiency, $f_{\rm i} = {\rm min}(2\epsilon_{\rm int}, 1)$, suggested by \citet{kruijssen12} who obtained estimates of the initial bound fraction from a series of small-scale star cluster formation simulations. However, this relation depends strongly on various properties of GMCs and on complex star formation and gas expulsion processes \citep[e.g.][]{adams00, baumgardt_kroupa07, goodwin09, smith_etal11, brinkmann_etal17, shukirgaliyev_etal18, farias_etal18}. It will be improved by future hydrodynamic simulations of turbulent clouds that self-consistently model star formation and stellar feedback (H.~Li et al. in prep.). For example, runs with $\epsff<1.0$ could be consistent with observations if a higher $f_{\rm i}$ was used for a given integrated star formation efficiency in the sub-grid model. At the same time, modifying this relation may also change the shape of the cluster mass function, since there is a strong correlation between the cluster particle mass and $\epsilon_{\rm int}$.

\begin{figure}
\includegraphics[width=\hsize]{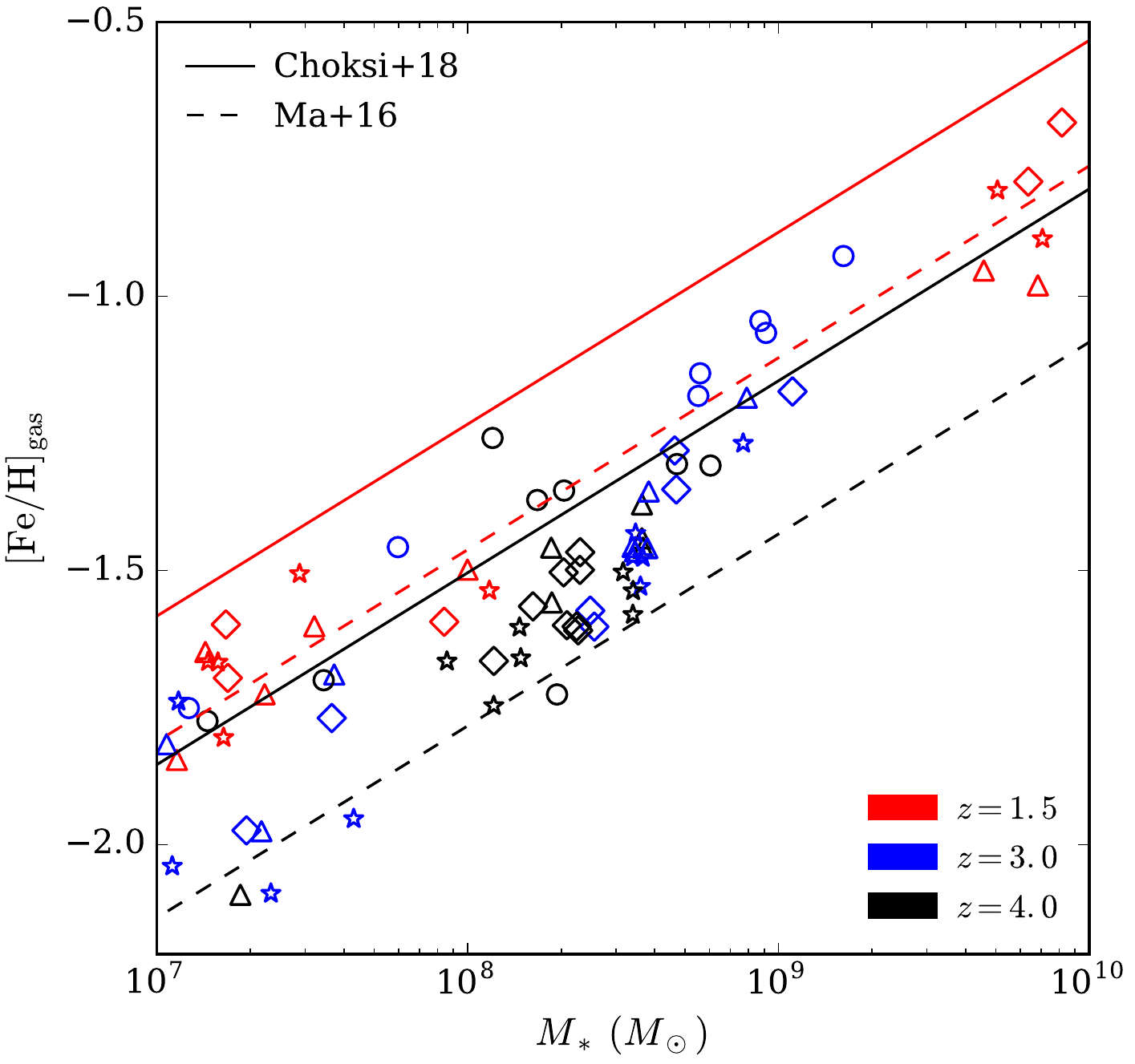}
\vspace{-4mm}
\caption{\small Gas-phase metallicity of galaxies with stellar mass $\Ms$ for four runs: SFEturb (circle), SFE50 (star), SFE100 (diamond), and SFE200 (triangle). We show results for three snapshots: at redshift 4 (black), 3 (blue), and 1.5 (red). Total stellar mass is obtained from stars within twice the stellar half-mass radius. Gas-phase metallicity is weighted by the molecular gas mass.
For reference, two analytical expressions for the evolution of gas-phase metallicity are shown as solid \citep{choksi_etal18} and dashed \citep{ma_etal16} lines for the limiting epochs, $z=4$ and $z=1.5$.
}\label{fig:gas_metal_history}
\end{figure}

\subsection{Galactic metallicity and GC metallicity}

Detailed simulations presented here can be used to calibrate simpler, semi-analytical models of GC formation. In a series of papers \citep{muratov_gnedin10,li_gnedin14,choksi_etal18, choksi_gnedin18} we have developed a model for the formation and evolution of massive star clusters that would evolve into GCs. This model takes dark matter halo merger trees from any cosmological simulation and supplements them with the prescriptions for galaxy stellar mass, cold gas mass, and metallicity based on the observed scaling relations. Star cluster formation is triggered at specific output epochs when the galaxy experiences faster mass growth than a chosen threshold. The model has only two adjustable parameters but makes a wide range of predictions for the mass, age, and metallicity distributions of surviving clusters that are in good agreement with the observations. 

In this model star clusters inherit the average metallicity of the gas in their host galaxy at the time of formation. Observations of the gas-phase metallicity for high-redshift galaxies are scarce \citep[e.g.][]{mannucci_etal09}, and therefore the model has to rely on uncertain extrapolation of the low-redshift relation. Also, the metallicity of low-redshift GCs is typically expressed as the iron abundance, which is not directly accessible in high-redshift galaxies and has to be inferred from the total metallicity. Here we use our simulation predictions for galaxy metallicity to test the accuracy of such extrapolations.

In \autoref{fig:gas_metal_history} we examine the gas-phase metallicity as a function of stellar mass of the main galaxy and its satellites at three different redshifts. The gas-phase iron metallicity, $\rm [Fe/H]_{gas}$, is defined as the molecular mass-weighted iron metallicity of all gas cells within half of the virial radius.
As the stellar-phase metallicity shown in \autoref{fig:metal_history}, the gas-phase metallicity increases with stellar mass and toward lower redshift. Both the mass- and redshift-dependence are in general agreement with those found in the FIRE simulations \citep{ma_etal16}.

The relation adopted in the model of \citet{choksi_etal18} appears to overestimate the gas-phase iron abundance by $\approx 0.2$~dex at $z=1.5$. At higher redshift the model is in closer agreement with the current simulations. Given the difficulty of accurate calibration of metallicity evolution and the wide scatter of values for different galaxies, we should take the 0.1-0.2~dex offset as a systematic uncertainty of the predictions of the analytical model for GC metallicity.

On the other hand, it is possible that our simulations underestimate the gas metallicity because the predicted metallicity distribution of surviving clusters does not extend to $\feh>-0.4$ to match the Galactic GCs. There are two possible ways to push the model metallicity distribution to higher values. The first is to form clusters at later times in more massive, more metal-rich galaxies. However, in order to reach $\feh\approx0$ according to the current MMR, GCs would need to form at $z<1$. The other way is to raise the normalization of the galactic MMR by about 0.2~dex, which would bring it into agreement with the analytical model. However, the current MMR in our simulations is a natural consequence of metal enrichment via large-scale galactic outflows due to efficient stellar feedback. Reducing the strength of feedback (e.g. run SFE50-SNR3) can increase the metallicity by suppressing outflows, but at the cost of higher star formation rate that is inconsistent with the abundance matching result.

\begin{figure}
\includegraphics[width=\hsize]{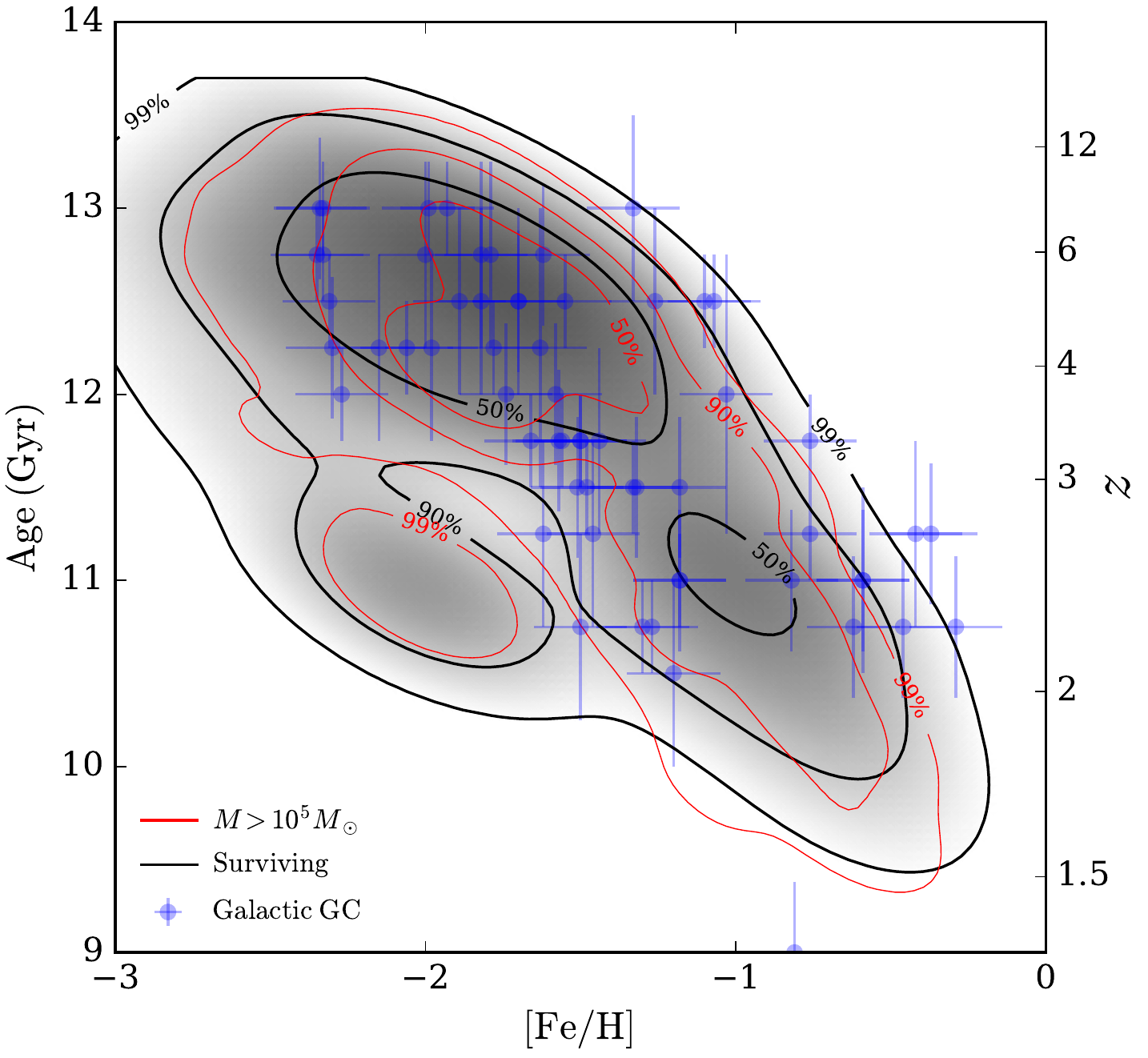}
\vspace{-4mm}
\caption{\small Age-metallicity distribution of model clusters in run SFE100 with initial mass higher than $10^5\Msun$ (red) and of surviving clusters with remaining mass higher than $10^4\Msun$ at $z=0$ (black). The percentage levels shown at each contour line represent the fraction of the number of model clusters enclosed within the corresponding contour. The observed age and metallicity of the Galactic GCs from \citet{leaman_etal13} are overplotted for comparison (blue circles with errorbars).
}\label{fig:age_met}
\end{figure}

\subsection{Age-metallicity distribution of surviving clusters}

An important relation that has emerged over the past few years in observations \citep[e.g.][]{dotter_etal10,leaman_etal13,vandenberg_etal13} and models \citep{muratov_gnedin10,li_gnedin14,choksi_etal18,kruijssen_etal18} is the age-metallicity relation for Galactic GCs. Metal-rich clusters form systematically later than metal-poor clusters, although the relation has a large scatter and significant irreducible uncertainty in measurements of the absolute ages. Such a relation is a robust prediction of the analytical models described above, because to achieve higher metallicity clusters need to form in more massive galaxies, and therefore, typically later. The shape of the age-metallicity relation has not noticeably changed from the earlier versions of the model \citep{muratov_gnedin10,li_gnedin14} to the later \citep{choksi_etal18}.

In \autoref{fig:age_met} we show the age-metallicity relation for surviving clusters formed in the current simulations. It is very close to the result of the analytical model (not shown for clarity). It also matches the observed distribution of ages and metallicities for the Galactic GCs.

We showed in \autoref{sec:results-massfunction} that majority of clusters with initial mass below $10^5\Msun$ disrupt completely, while majority of clusters with initial mass above $10^5\Msun$ survive until the present. This mass scale may be used as a simple selection criterion of GC candidates, without having to calculate the detailed dynamical evolution. We compare the contours enclosing 50\%, 90\%, and 99\% of the actual surviving clusters (shown in black) with those of clusters with the initial mass above $10^5\Msun$ (shown in red). As expected, the two sets of contours are similar, with only slight differences at lowest metallicity, where for the actual surviving clusters are older by $0.2-0.3$~Gyr.

\section{Summary} \label{sec:summary}

We have developed an algorithm to calculate the strength of tidal field experienced by star cluster particles in cosmological simulations. The very high spatial resolution of our adaptive-mesh simulations, $3-6$~pc, allows accurate calculation of the tidal field in realistic galactic environments. Using the eigenvalues of the tidal tensor along cluster trajectory, we calculate the rate of cluster disruption in simulation runtime. We quantify the evolution of the cluster mass function and metallicity distribution across cosmic time, and determine a set of clusters that remain gravitationally bound until the present. Our main conclusions are summarized below:
\begin{itemize}
\item The strength of the tidal field decreases as clusters migrate away from dense central parts of the host galaxy. In contrast to the result of \citet{renaud_etal17}, tidal disruption of clusters in our simulations is most important during the first Gyr after formation. Having high enough spatial resolution of the simulation is crucial for accurate determination of the tidal field.

\item As a result of tidal disruption, over 70\%-80\% of all clusters with initial mass above $10^5\Msun$ have remaining mass smaller than $10^4\Msun$ at present. Essentially all clusters with initial mass below $10^5\Msun$ are disrupted before the present in all the runs.

\item The fraction of stellar mass contained in bound star clusters decreases as the simulations approach lower redshift. The fraction evolves from 6.2\% at $z=4$ to 1.4-1.8\% at $z=1.5$ for SFE100 run. Extrapolating to $z=0$, where the expected stellar mass of a Milky Way-sized galaxy is $6\times10^{10}\Msun$, this fraction drops to $\sim 4\times10^{-4}$. The total mass of surviving clusters at \mbox{$z=0$} increases with $\epsff$ adopted in the simulations: $(1-1.5)\times10^7\Msun$ for SFE50 run, $(1.8-2.4)\times10^7\Msun$ for SFE100 run, and $(4.8-6)\times10^7\Msun$ for SFE200 run. The lower and upper limits are obtained from the tidal disruption calculations using $\lambda_m$ and $\lambda_1$, respectively. Run SFE100 is most consistent with the observed mass of the GC system in Milky Way-sized galaxies \citep[e.g.][]{spitler_forbes09}.

\item The shape of the mass function of bound clusters evolves from an initial truncated power-law to a peaked distribution at $z=1.5$. Extrapolating the disruption to $z=0$ results in too many low-mass clusters compared to the observed Galactic GCs.  This discrepancy may be due a necessarily simplified way of calculating tidal disruption after the last available simulation output or a lack of mass-loss implementation by tidal shocks.

\item We find a scaling relation between the average metallicity of young clusters and stellar mass of their host galaxy. This stellar mass-metallicity relation is consistent with observations of galaxies at $z=1-4$ and with results of other comparable cosmological simulations \citep{ma_etal16}.

\item We find that the metallicity distribution of surviving GCs spans a wide range $-2.8 < \feh < -0.5$ and is generally similar to the observations of Galactic GCs. The model produces more metal-poor clusters and not enough very metal-rich clusters to fully match the observations.

\item Massive, low-metallicity clusters acquire an internal spread of $\sigma_\feh \approx 0.1-0.2$, with the largest values up to 0.4~dex for clusters with $\rm [Fe/H]<-2.5$~dex. The iron spread due to gas accretion during the early stage of cluster formation can account for the observed iron spread in Galactic GCs, except for a few GCs with $\sigma_\feh \gtrapprox0.2$~dex, which are suspected to be the debris of disrupted dwarf galaxies.

\item GCs surviving until the present exhibit a clear age-metallicity relation, where metal-rich clusters are systematically younger than metal-poor clusters by up to 3~Gyr. This relation is consistent with the predictions of the semi-analytical model of GC formation \citep{li_gnedin14,choksi_etal18}.
\end{itemize}

\section*{Acknowledgements}

We thank Nick Gnedin for continuous help with the ART code, and Eric Bell, Mark Gieles, Gillen Brown, and Xi Meng for helpful discussions.
We thank the anonymous referee for detailed comments and suggestions.
We also thank Michigan Data Science Team (MDST) for providing us computing resources in the high-performance computing centre Flux, which was used to perform most of the simulations in this paper. MDST data computation is supported by Advanced Research Computing - Technology Services at the University of Michigan and NVIDIA.
This work was supported in part by NSF through grant 1412144.

%%%%%%%%%%%%%%%%%%%%%%%%%%%%%%%%%%%%%%%%%%%%%%%%%%



\bibliographystyle{mnras}
\bibliography{references} % if your bibtex file is called example.bib

% Don't change these lines
\bsp	% typesetting comment
\label{lastpage}
\end{document}